\documentclass[journal]{IEEEtran}
\usepackage{amsmath,amsfonts}
\usepackage{amsthm}
\usepackage{amssymb}
\usepackage{csquotes}
\usepackage{blkarray}
\usepackage{bm}
\usepackage{cases}
\usepackage{xcolor}
\usepackage{hyperref}
\usepackage{booktabs}
\usepackage{multirow}
\usepackage{setspace}

\usepackage{algorithm}
\usepackage[noend]{algpseudocode}

\usepackage{array}
\usepackage[caption=false,font=normalsize,labelfont=sf,textfont=sf]{subfig}
\usepackage{textcomp}
\usepackage{stfloats}
\usepackage{url}
\usepackage{verbatim}
\usepackage{comment}
\usepackage{graphicx}
\newtheorem{theorem}{Theorem}

\newtheorem{remark}{Remark}
\newtheorem{lemma}{Lemma}
\newtheorem{definition}{Definition}

\newtheorem{observation}{Observation}
\usepackage{caption}
\usepackage[noadjust]{cite}

\usepackage{float}
\floatstyle{ruled}
\newfloat{model}{thp}{lop}
\floatname{model}{Model}

\definecolor{niceblue}{rgb}{0, 0.5, 1.0}
\definecolor{niceblue}{rgb}{0.125, 0.406, 0.852}

\hypersetup{
    colorlinks=true,
    linkcolor=niceblue,
    filecolor=magenta,      
    urlcolor=niceblue,
    citecolor=niceblue}
\newcommand\undermat[2]{%
  \makebox[0pt][l]{$\smash{\underbrace{\phantom{%
    \begin{matrix}#2\end{matrix}}}_{\text{$#1$}}}$}#2}

\usepackage{balance}

\renewcommand{\baselinestretch}{0.95}

\begin{document}
% \title{Tighter Global Performance Guarantees for Learned Piecewise Linear Power Flow Models}
\title{Global Performance Guarantees for Neural\\ Network Models of AC Power Flow}
%\title{Global Performance Guarantees for Piecewise Linear Power Flow Models Learned from Data}
\author{Samuel Chevalier,~{\textit{Member, IEEE}}, and Spyros Chatzivasileiadis,~{\textit{Senior Member, IEEE}}
%\thanks{Samuel Chevalier is supported by the HORIZON-MSCA-2021 Postdoctoral Fellowship Program, Project \#101066991 – TRUST-ML. Spyros Chatzivasileiadis is supported by the ERC Starting Grant VeriPhIED, Grant Agreement No. 949899. Both authors are with the Department of Wind and Energy Systems at the Technical University of Denmark (DTU), Kongens Lyngby, Denmark. Emails: \{schev,spchatz\}@dtu.dk.}}
}

%\markboth{Journal of \LaTeX\ Class Files,~Vol.~18, No.~9, September~2020}%
%{How to Use the IEEEtran \LaTeX \ Templates}

\maketitle

\begin{abstract}
Machine learning, which can generate extremely fast and highly accurate black-box surrogate models, is increasingly
being applied to a variety of AC power flow problems. Rigorously verifying the accuracy of the resulting black-box models, however, is computationally challenging. This paper develops a tractable neural network verification procedure which incorporates the ground truth of the non-linear AC power flow equations to determine worst-case neural network prediction error. Our approach, termed Sequential Targeted Tightening (STT), leverages a loosely convexified reformulation of the original verification problem, which is an intractable mixed integer quadratic program (MIQP). Using the sequential addition of targeted cuts, we iteratively tighten our formulation until either the solution is sufficiently tight or a satisfactory performance guarantee has been generated. After learning neural network models of the 14, 57, 118, and 200-bus PGLib test cases, we compare the performance guarantees generated by our STT procedure with ones generated by a state-of-the-art MIQP solver, Gurobi 11.0. We show that STT often generates performance guarantees which are far tighter than the MIQP upper bound. 

\end{abstract}


\section{Introduction}
\IEEEPARstart{M}{achine} Learning (ML) and Artificial Intelligence (AI) are set to profoundly disrupt many facets of the power and energy sectors 
%in the coming decades 
according to utility executives~\cite{AI_insights}. To proactively anticipate the massive rollout of ML-aided technology,
%across many different industries,
the push for so-called Trustworthy AI is being embraced by many governmental organizations. In April of 2021, the EU Commission proposed the creation of regulations which will assess AI technologies across multiple safety and compliance standards~\cite{kop2021eu}; high-risk applications will face market entrance and certification requirements. %prior to deployment~\cite{kop2021eu}.
Similarly, the National Institute of Standard and Technologies (NIST) in the US is developing trustworthy AI technical standards related to 
%across metrics including
validation, verification, and interpretability~\cite{NIST_AI,national2019us}.

Motivated by these forthcoming regulations, this paper focuses on the problem known as performance verification, where the output prediction of a ML model is systematically compared to the ground truth in order to quantify model performance. By assessing accuracy across the full input domain, model performance can be verified as satisfactory or not; depending on these results, the model can then either be deployed for application or sent back for retraining. Verification problems are often tackled by formulating an optimization problem where the neural network (NN) under examination is transformed into an equivalent set of optimization constraints.
%by transforming a neural network (NN) model into an equivalent set of optimization-based constraints. 
Popular verification methodologies have utilized, e.g., convex relaxations of activation functions, semidefinite programming (SDP) relaxations, Lagrangian relaxations, mixed-integer reformulations, and various norm-bounding approaches~\cite{Wong:2018, Krishnamurthy:2018, Dathathri:2020, Fazlyab:2022, Dvijotham:2020, Tjeng:2017}. The most promising approaches recently competed at VNN-COMP-22~\cite{brix2023fourth}, a NN verification competition; $\alpha,\beta$-CROWN~\cite{Wang:2021} was the top performing algorithm in the last three competitions.

% Salman:2019

% \footnote{The $\alpha,\beta$-CROWN algorithm recently won the 2022 competition as well.}.

In the field of power systems, the AC power flow equations are foundationally important for any problem related to network optimization, dynamics, or control. Given ML's ability to efficiently learn complex relationships in high dimensional spaces, the problem of learning surrogate power flow mappings has become a very relevant research direction in recent years. Researchers from RTE utilized graph neural networks to quickly estimate network power flow solutions~\cite{donon2020neural}; they also exploited guided dropout in~\cite{2018arXivDonnot} to predict power flow solutions across exhaustive N-1 and N-2 contingencies. The bilinear nature of power flow was exploited in~\cite{hu2020physics} to train a NN which generalizes effectively. Authors in~\cite{Kody:2021} learned a piecewise linear power flow mapping which was then embedded in the AC unit commitment (UC) problem. 
%Learning power flows in distribution networks was explored in~\cite{Park:2021}. 
On top of these works sits a mountain of emerging research which uses ML to solve or estimate \textit{optimal} power flow (OPF) type problems~\cite{Zamzam:2020,Hasan:2020,NELLIKKATH2022108412, Ilgiz_IREP}. Rigorously verifying NN performance is a relevant problem for any of the works listed in this paragraph.

%donon2019graph

In this paper, we seek to verify the performance of a NN which has learned power flow physics. We expect that this contribution can be the cornerstone for the verification of a wide family of ML algorithms, e.g., ones that predict power system states or determine control set-points.
%We expect that such a work can constitute the cornerstone for the verification of a wide family of ML algorithms, e.g., ones that aim to predict the power system state (i.e., state estimation) or determine control set-points. 
Similar to the MIQP posed in~\cite{eydenberg4245628physics}, we verify performance by comparing the NN output directly to the ground-truth power flow (${\rm pf}$) equations and determining the maximum error: $\max_x\{{\rm pf}(x)-{\rm NN}(x)\}$. Thus, in the verification problem, we exploit the fact that we have direct access to the physics that the ML model is attempting to learn. While advancing quickly, state-of-the-art verification methods from the ML community~\cite{brix2023fourth} cannot be directly applied to this problem for three primary reasons. 
\begin{itemize}

\item First, verification problems which include quadratic or bilinear nonconvexities (e.g., from power flow) tend to be ``incomplete" verification problems. That is, most state-of-the-art solvers~\cite{Wang:2021} cannot provide conclusive verification decisions on such problems if the bound propagation routines are inconclusive.

\item Second, we seek to verify NN performance across a global set of inputs (i.e., any set of feasible network loadings). Many of the methods participating in~\cite{brix2023fourth} are designed around the adversarial robustness problem, where inputs are perturbed by a very small epsilon. In that case, norm bound propagation approaches (which $\alpha,\beta$-CROWN is built on) are very successful. However, verifying a model across the full input range (which we seek to do, in order to provide global performance guarantees to system operators) requires a more general error tightening approach.

\item Finally, existing verification tools are designed around solving binary verification problems (i.e., safe vs unsafe); they don't return maximal error guarantees or safety margins, which are important in safety critical applications.

\end{itemize}

% The problem of learning power flow mappings has become a very relevant research direction in recent years. Researchers from RTE utilized graph neural networks to quickly estimate network power flow solutions~\cite{donon2019graph,donon2020neural}. They also exploited guided dropout in~\cite{2018arXivDonnot} to predict power flow solutions across exhaustive N-1 and N-2 contingencies. The bilinear nature of power flow was exploited in~\cite{hu2020physics} to train a NN which generalizes well. Authors in~\cite{Kody:2021} learned a piecewise linear power flow mapping which was then embedded in the AC unit commitment (UC) problem. Learning power flows in distribution networks was explored in~\cite{Park:2021}. On top of this work sits a mountain of emerging research which uses ML to solve or estimate optimal power flow (OPF) type problems~\cite{Zamzam:2020,Hasan:2020,NELLIKKATH2022108412}. Rigorously verifying NN performance is a relevant problem for any of the works listed in this paragraph.

Within the power systems community, the problem of NN performance verification was first proposed in~\cite{Venzke:2020_fg}, where NNs were reformulated into Mixed Integer Linear Programs (MILPs). After training NNs to classify stable and unstable operating points, these MILPs were solved to ($i$) rigorously verify the NN estimated stability regions, and ($ii$) discover adversarial examples. This approach was extended to regression NNs in~\cite{Venzke:2020}. Using the DC-OPF as the guiding application, it was possible for the authors to include the ground truth inside the MILP for NN verification. After training NNs to solve the DC-OPF problem, the MILP reformulation was utilized to determine both maximum constraint violations and maximum solution sub-optimality (relative to the true DC-OPF solution).

%``($i$) maximum constraint violations, ($ii$) maximum distance between predicted and optimal decision variables, and ($iii$) maximum sub-optimality (relative to the true DC-OPF solution)" 

These methods were recently extended to the AC-OPF problem in~\cite{NELLIKKATH2022108412}. Contrary to the DC-OPF, which is a linear program and does not change the nature of the MILPs when inserted in the verification problem, AC-OPF is non-linear: this turns the MILP into a mixed integer \textit{quadratically constrained} program (MIQCP). In trying to assess worst case line flow violation on the 39-bus system, authors stated the following: \vspace{0.1cm}
\begin{displayquote}
\begin{spacing}{0.89}
\textit{\small
``we were unable to compute the worst-case line flow constraint
violation since the MIQCQP problem could not be solved to zero optimality gap within 5 hr. This highlights the computational challenges
associated with the extraction of the worst-case guarantees for AC-OPF..."}
\end{spacing}
\end{displayquote}\vspace{-0.25cm}

In order to tackle this challenge,  
%Our paper does not tackle this OPF problem directly. 
we step back from the OPF problem and more broadly target ($i$) the nonlinear power flow equations and ($ii$) the mixed integer nature of ReLU-based NN networks which, in combination, are at the heart of the verification challenge. Our ability to efficiently solve this problem will remove the barriers associated with solving NN verification problems for any process governed by quadratic equations (including the AC-OPF, AC Optimal Control, and many other problems, extending even beyond power systems). 
%So far, to the best of our knowledge, there has been no attempt to tackle this challenge in the power systems literature. 
Our approach utilizes a common convexified (SDP) framework for relaxing both the NN binaries and the quadratic power flow variables. Since the resulting formulation tends to be very loose, we tighten the solutions with a combination of equality constraints proposed in recent works~\cite{Dvijotham:2020} and the sequential addition of Sherali-Adams cuts (also known as Reformulation-Linearization Technique cuts); such cuts have been recently utilized for NN verification within the ML community~\cite{Ma:2020,Lan:2022}, tightening SDP relaxations of the UC problem~\cite{Fattahi:2017}, and Optimization Based Bound Tightening (OBBT)~\cite{CENGIL2022108275}.

Given the challenging nature of NN verification in the context of power flow emulation and the lack of existing solution techniques, our paper contributions follow:

\begin{enumerate}

    \item We develop a verification framework, termed Sequential Targeted Tightening (STT), which efficiently generates global performance guarantees for any NN verification problem which can be posed as a MIQP.
    
    \item Through the sequential addition of Sherali-Adams cuts, we iteratively generate tighter and tighter upper bounds on model performance in a computationally expedient manner, thus enabling faster verification solutions.
    
    % thus allowing the procedure to terminate whenever a performance metric has been achieved. 
    
    \item We apply STT to the problem of verifying power flow predictions, and we demonstrate how STT bounds can significantly outperform the bounds generated by a state-of-the-art MIQP solver (Gurobi 11.0).

    %develop a verification framework which helps remove the barriers associated with deplo
    
    %\item We define a NN verification problem which simultaneously relaxes both the mixed integer reformulation of the NN and the quadratic power flow equations into a common semidefinite programming space.
    
    %\item In order to tighten the convexified solution, we generate an exhaustive ($n^2$) set of Sherali-Adams cuts which project linear bounding constraints into a lifted space.
    
    %\item Through a procedure we term Sequential Targeted Tightening (STT), we selectively add subsets of the Sherali-Adams cuts back into the convex verification problem, thus tightening the worst case performance bound.
    
\end{enumerate}

%The remainder of this paper is structured as follows. 
In Sec. \ref{sec:NN_performance}, we pose the challenging NN performance verification problem. We then tighten its convexified counterpart using Sherali-Adams cuts and a novel sequential targeted tightening (STT) procedure. In Sec. \ref{sec:pf_application}, we apply this STT approach to the power flow verification problem. In Sec. \ref{sec:test_setup}, we set up the testing procedures (data collection and NN training) we use to assess STT. Test results are presented in Sec. \ref{sec:results}, and we conclude in Sec. \ref{sec:conclusion}. %where we offer highlights and limitations of the proposed methodology.% along with an outline for next steps and future work.

\section{Neural Network Performance Assessment}\label{sec:NN_performance}
In this section, we introduce the problem of globally verifying NN performance against the underlying ground truth that the NN emulates. In this paper, a NN consists of linear transformations and ReLU activation functions; this mapping can be treated as a piecewise linear model through an exact reformulation (Appendix~\ref{AppA}). In this section, we consider a NN which acts as a piecewise linear surrogate model that captures feasibility constraints (e.g., power flow). As posed, the associated verification problem is intractably complex. Accordingly, we present sequential relaxations and targeted tightenings which allow us to compute fairly tight upper bounds on the verification problem using a convex SDP solver.

\subsection{Performance verification of general surrogate constraints}
\begin{figure}
\includegraphics[width=\columnwidth]{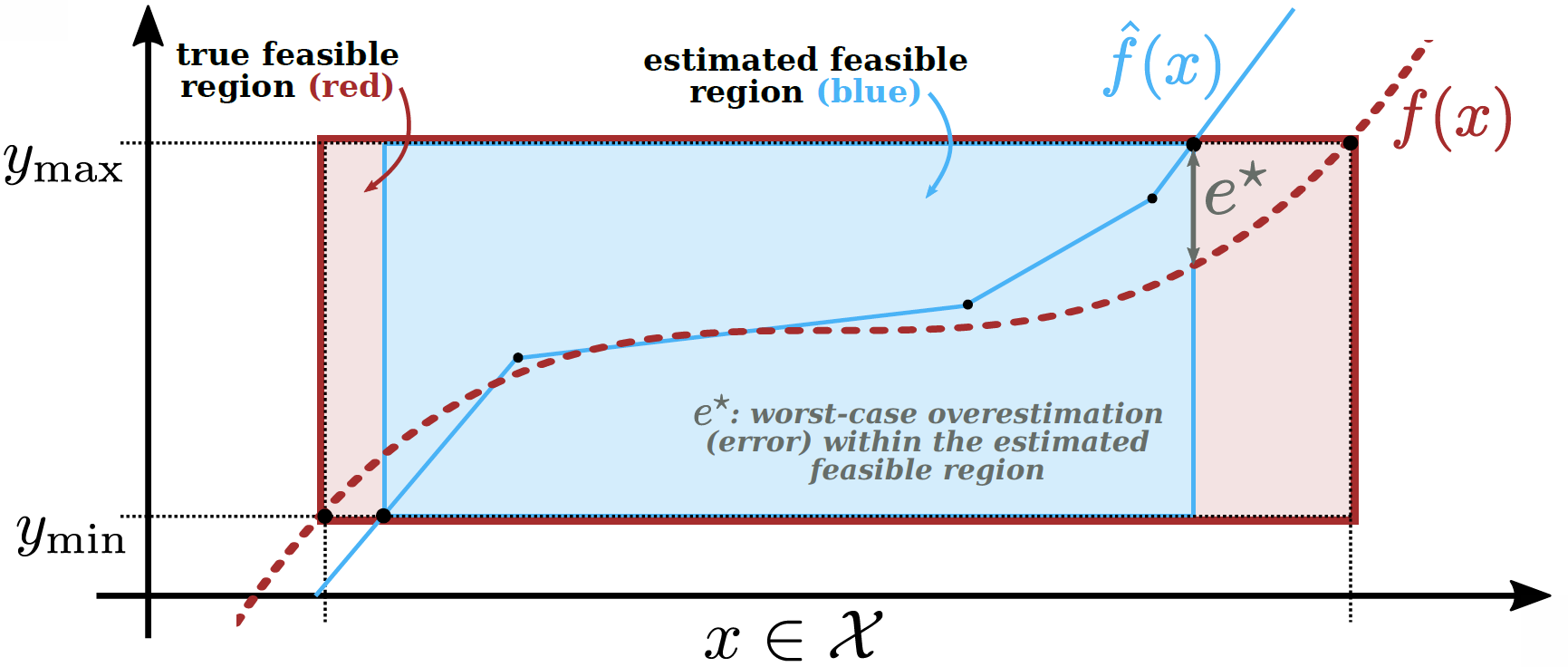}
\caption{The value $e^{\star}$ represents the worst-case overestimation of the surrogate model ${\hat f}(x)$, with respect to the ground truth function $f(x)$, within the feasible space estimated by the surrogate: $y_{\rm min}\le{\hat f}(x)\le y_{\rm max}$.}
\label{fig:error}
\end{figure}

We motivate the class of problems addressed in this paper by considering the following optimization problem:
\begin{subequations}\label{eq: opt_problem}
\begin{align}
c^{\star}=\min_{x\in {\mathcal X}_f}\;\, & c(x)\\
{\rm s.t.}\;\, & f(x)-p_c\ge 0,\label{eq: original_constraint}
\end{align}
\end{subequations}
where $c(x)$, $x\in {\mathbb R}^n$, is some cost function, $f(x)$ represents a series of challenging optimization constraints, and $p_c\in {\mathbb R}^m$ is a vector of constant parameters. Due to the computational challenge associated with solving \eqref{eq: opt_problem}, recent approaches have sought to learn high-fidelity surrogate models, ${\hat f}(x)$, which effectively replace $f(x)$ in \eqref{eq: original_constraint}. The resulting solution, $\hat{c}^{\star}$, can be a good approximation of the true cost:
\begin{subequations}\label{eq: }
\begin{align}
c^{\star}\approx \hat{c}^{\star}=\min_{x\in {\mathcal X}_f}\;\, & c(x)\\
{\rm s.t.}\;\, & \hat{f}(x)-p_c\ge0.\label{eq: surrogate_costraint_OG}
\end{align}
\end{subequations}
In this paper, we study the important problem of rigorously bounding the error associated with the learned surrogate model. To do so, we seek to solve the optimization problem
\begin{subequations}\label{eq: verification}
\begin{align}e^{\star}_i = \max_{x\in {\mathcal X}_f}\;\; & f_{i}(x)-\hat{f}_{i}(x)\label{eq: diff_max}\\
{\rm s.t.}\;\; & \hat{f}(x)-p\ge0.\label{eq: surrogate_costraint}
\end{align}
\end{subequations}
The objective function in \eqref{eq: diff_max} maximizes the difference between the surrogate model's $i^{\rm th}$ output prediction and the original equation it emulates. Thus, $e^{\star}_i$ captures the NN underestimation \textit{error}; overestimation can be captured by maximizing the negative of \eqref{eq: diff_max}. Constraint \eqref{eq: surrogate_costraint} ensures that the model stays within the same feasible region utilized in its deployment (i.e., \eqref{eq: surrogate_costraint_OG}). However, we parameterize this feasible region with vector $p$, instead of $p_c$, where $p$ corresponds to the largest feasible region in which the model may be deployed (e.g., different loading scenarios in a power system)\footnote{As an example, a constraint $f_i(x)$ may be bounded by $0\le f_{i}(x)\le1$ or $1\le f_{i}(x)\le2$ in \eqref{eq: opt_problem}. Therefore, in verification, the values of $p$ are set to implement the broader bound of $0\le f_{i}(x)\le2$.}. We therefore call $e_i^{\star}$ a \textit{performance guarantee}, since it represents the maximum possible amount of error between the surrogate function and the ground truth. Fig.~\ref{fig:error} depicts the worst-case error associated with a function $f(x)$ and its piece-wise linear surrogate ${\hat f}(x)$. Notably, the error is evaluated within the region that the surrogate function predicts to be feasible.

In this paper, we restrict $f(x)$ to the set of multivariate quadratic
%\footnote{Linear terms can also be accommodated in \eqref{eq: quad_const} without complication.} 
functions\footnote{Any set of polynomials can be reformulated into this quadratic form~\cite{MOSEK}.}   given by
%This paper restricts $f(x)$ to multivariate quadratic functions:
%\footnote{Linear terms can also be accommodated in \eqref{eq: quad_const} without complication.} 
\begin{align}\label{eq: quad_const}
    f_{i}(x)=x^{T}\bm{M}_{i}x=\sum_{j,k}m_{i,j,k}x_{j}x_{k}.
\end{align}
We also restrict the surrogate functions ${\hat f}(x)$ to ones which can be constructed with ReLU activation functions and linear transformations, as in the following neural network (NN):
\begin{align}\label{eq: NN_ReLU}
y=\hat{f}(x)\triangleq \bm{W}^{(n)}\sigma(\cdots\sigma(\bm{W}^{(0)}x+b^{(0)})\cdots)+b^{(n)},
\end{align}
where $\sigma(x) \triangleq \max({x,0})$, and the parenthetical superscript notation, e.g., $b^{(i)}$, indicates an item associated with the $i^{\rm th}$ layer of a NN. Using the MILP reformulation provided in Appendix \ref{AppA} and a therein defined selection vector $s_i$ which selects the $i^{\rm th}$ NN output, the verification problem \eqref{eq: verification} can be written as a non-convex mixed integer quadratic program (MIQP). This MIQP is stated in Model \ref{model:MIQP}, where $u\in {\mathbb R}^{t}$, defined in Appendix \ref{AppA}, consists of the NN input $x$, the binary vector $\beta \in \{0,1\}^d$, and any number of intermediate NN states $\hat x$. For specificity, $\bm{A}_{x}x-b_{x}\ge0$ has replaced the constraints on the input variable previously given by $x\in\mathcal{X}_{f}$.

\begin{model}[h]
\caption{\hspace{-0.1cm}\textbf{:} NN Performance Assessment via MIQP}
\label{model:MIQP}
\vspace{-0.25cm}
\begin{align}e_{i}^{{\star}}=\max_{x,{\hat{x}},\beta}\;\; & x^{T}\bm{M}_{i}x-s_{i}^{T}(\bm{W}^{(n)}\hat{x}^{(n)}+b^{(n)})\nonumber\\
{\rm s.t.}\;\; & \bm{A}_{x}x-b_{x}\ge0\tag{Input Bound}\label{eq: MIQPb}\\
 & \bm{A}_{{\rm nn}}u-b_{{\rm nn}}\ge0\label{eq: MIQPc}\tag{NN Mapping}\\
 & (\bm{W}^{(n)}\hat{x}^{(n)}+b^{(n)})-p\ge0\label{eq: MIQPd}\tag{Output Bound}\\
 & \beta\in\{0,1\}^{d}.\label{eq: MIQPe}\tag{Binary Vector}
\end{align}
\vspace{-0.5cm}
\end{model}
Notably, the NN output vector is not explicitly defined in Model \ref{model:MIQP}, but its $i^{\rm th}$ value is explicitly selected in the objective function via ${\hat f}_i=s_{i}^T(\bm{W}^{(n)}\hat{x}^{(n)}+b^{(n)})$.

\subsection{Semidefinite relaxations of the binary and quadratic terms}
Despite the significant advances made by commercial MIQP solvers like Gurobi in recent years, Model \ref{model:MIQP} is still very difficult to solve in practice. To exploit conic programming, whose problems are much easier to solve, Model \ref{model:MIQP} can be relaxed into a semidefinite program (SDP) through simultaneous relaxations of the binary constraints and quadratic objective terms. First, we define the matrices
\begin{align}
\mathcal{X} \triangleq\left[\begin{array}{cc}
1 & x^{T}\\
x & \bm{X}
\end{array}\right],\quad\; \mathcal{B} \triangleq\left[\begin{array}{cc}
1 & \beta^{T}\\
\beta & \bm{B}
\end{array}\right].
\end{align}

%\begin{align}
%\mathcal{X} & \triangleq\left[\begin{array}{cc}
%1 & x^{T}\\
%x & \bm{X}
%\end{array}\right]\\
%\mathcal{B} & \triangleq\left[\begin{array}{cc}
%1 & \beta^{T}\\
%\beta & \bm{B}
%\end{array}\right].
%\end{align}

\begin{remark}
Matrices $\mathcal{X}$ and $\mathcal{B}$ have the following properties:
\begin{itemize}
\item If ${\rm rank}(\mathcal{X})=1$, then $\bm{X}=xx^{T}$. 
\item If ${\rm rank}(\mathcal{B})=1$ and ${\rm diag}(\mathcal{B})=\beta$, then $\bm{B}=\beta\beta^{T}$ and $\beta$ is a vector of binaries: $\beta\in\{0,1\}^d$~\cite{MOSEK}.
\end{itemize}
\end{remark}
Since rank constraints are non-convex, we relax them both in order to obtain the convex SDP stated in Model \ref{model:SDP}.
\begin{model}[h]
\caption{\hspace{-0.1cm}\textbf{:} NN Performance Assessment via SDP}
\label{model:SDP}
\vspace{-0.25cm}
\begin{align}
{\tilde e}_{i}^{{\star}}=\max_{\mathcal{X},\mathcal{B},\hat{x}}\;\; & {\rm tr}(\bm{M}_{i}\bm{X})-s_{i}^{T}(\bm{W}^{(n)}\hat{x}^{(n)}+b^{(n)})\nonumber\\
{\rm s.t.}\;\; & \bm{A}_{x}x-b_{x}\ge0\tag{Input Bound}\\
 & \bm{A}_{{\rm nn}}u - b_{{\rm nn}}\ge 0 \tag{NN Mapping}\\
 & \bm{W}^{(n)}\hat{x}^{(n)}+b^{(n)}-p\ge0\tag{Output Bound}\\
 & \mathcal{X}\succeq0,\;\mathcal{X}_{1,1}=1\tag{Input PSD Cone}\\
 & \mathcal{B}\succeq0,\;\mathcal{B}_{1,1}=1\tag{Binary PSD Cone}\\
 & 0\le{\rm diag}(\bm{B})=\beta\le 1.\tag{Binary Tightening}
\end{align}
\vspace{-0.5cm}
\end{model}

As noted in~\cite{MOSEK}, additional constraints \eqref{eq: bin_C1}-\eqref{eq: bin_C4} can be added to help tighten the binary matrix solution:
\begin{subequations}\label{eq: binary_SAcuts}
\begin{align}
     0\le&\bm{B}_{i,j}\le1,\hspace{-40pt}&&\quad\forall i,j\label{eq: bin_C1}\\
 \bm{B}_{i,j}\le&\bm{B}_{i,i},\hspace{-40pt}&&\quad\forall i,j\label{eq: bin_C2}\\
 \bm{B}_{i,j}\le&\bm{B}_{j,j},\hspace{-40pt}&&\quad\forall i,j\label{eq: bin_C3}\\
 \bm{B}_{i,j}\ge&\bm{B}_{j,j}+\bm{B}_{i,i}-1,\hspace{-40pt}&&\quad\forall i,j.\label{eq: bin_C4}
\end{align}
\end{subequations}
However, \eqref{eq: bin_C1}-\eqref{eq: bin_C4} represent approximately $4n^2$ constraints, most of which are automatically satisfied by the binary PSD cone constraint, so their direct inclusion quickly becomes intractable (even for, say, 100 binaries).

\label{eq: SDP1}

As it exists, the SDP in Model \ref{model:SDP} has two major issues. First, it is generally dual infeasible (i.e., unbounded), since the matrix $\bm X$ in the objective is unconstrained. This can be fixed by extending the constraints on input vector $x\in\mathcal{X}_{f}$ to the lifted semidefinite matrix $\bm X$. However, even with these additional constraints, there is a more central second issue: Model \ref{model:SDP} is generally a very loose relaxation, with ${\tilde e}_{i}^{{\star}} \gg {e}_{i}^{{\star}}$. This is primarily because the input constraints on $x$ are not strong enough to suppress the rank of $\bm X$, and the binary relaxation tends to be fairly loose. In the following subsection, we tighten the relaxation via Sherali-Adams cuts.

\subsection{Tightening loose verification SDP via Sherali-Adams cuts}
In its most basic application, the Sherali-Adams method tightens relaxations by taking the products of valid constraints~\cite{Sherali:2013}; it then linearizes the resulting products by replacing, e.g., quadratic terms $x_i*x_j$ with their lifted SDP matrix counterparts, (i.e., $\bm X_{i,j}$). In the case of linear constraints like ${\bm A}x\ge0$, valid Sherali-Adams cuts can be exhaustively added by taking an outer product of this expression with itself: ${\bm A}x({\bm A}x)^{T}\ge\bm{0}\Rightarrow{\bm A}\bm{X}{\bm A}^{T}\ge\bm{0}$~\cite{Fattahi:2017}. Notably, this procedure squares the total number of constraints (i.e., from $k$ linear constraints in ${\bm A}x\ge0$ to $k^2$ lifted constraints in $\textbf{A}\bm{X}\textbf{A}^{T}\ge\bm{0}$). Sherali-Adams, also known as the Reformulation-Linearization Technique (RLT)~\cite{Ma:2020,Lan:2022,Gopinath:2020}, is known as a ``lift and project" method, and it is closely related to the Lasserre hierarchies that have been successfully utilized in power system optimization in recent years~\cite{Molzahn:2015}.

To apply this procedure to Model \ref{model:SDP}, we first put it into a more canonical form. In doing so, we remark that four of the binary constraints implicitly result from Sherali-Adams cuts.  

\begin{remark}
Constraints \eqref{eq: bin_C1}-\eqref{eq: bin_C4} can be replaced by applying exhaustive Sherali-Adams cuts to constraint $0\le\beta\le1$.
\end{remark}
Accordingly, we capture $0\le\beta\le1$ in the canonical fashion of $\bm{A}_{\beta}\beta-b_{\beta}\ge0$, and then we aggregate all linear constraints into $\bm{A}u-b\ge0$, where $\bm{A}\in {\mathbb R}^{N \times M}$:
\begin{subnumcases}
{\bm{A}u-b\ge0\;\;\Leftrightarrow\;\;}
\bm{A}_{x}x-b_{x} \ge0\label{eq: Aub1}\\
\bm{A}_{\beta}\beta-b_{\beta} \ge0\label{eq: Aub2}\\
\bm{A}_{{\rm nn}}u-b_{{\rm nn}} \ge0\label{eq: Aub3}\\
\bm{W}^{(n)}\hat{x}^{(n)}+b^{(n)}-p \ge0\label{eq: Aub4},
\end{subnumcases}
where \eqref{eq: Aub1} bounds the input, \eqref{eq: Aub2} bounds the binaries, \eqref{eq: Aub3} captures the NN mapping constraints, and \eqref{eq: Aub4} bounds the output to the functionally feasible region. Applying exhaustive Sherali-Adams cuts to $\bm{A}u-b\ge0$ yields the outer product
\begin{equation}
\begin{aligned}
(\bm{A}u\!-\!b)(\bm{A}u\!-\!b)^{T} & \!=\!\bm{A}uu^{T}\!\bm{A}^{T}\!\!-\!\bm{A}ub^{T}\!\!-\!bu^{T}\!\bm{A}^{T}\!\!+\!bb^{T}\label{eq: Aub_outer}\\
&\ge\bm{0}.
\end{aligned}
\end{equation}
The $uu^{T}$ term in \eqref{eq: Aub_outer} will require us to expand the number of semidefinite variables. Accordingly, we define $\bm{\Gamma}$ as the lifted matrix variable for $uu^T$, and we define $\hat{\bm{\Gamma}}$ similarly:
%\begin{comment}
%\begin{align}\label{eq: Gamma}
%\bm{\Gamma}
% & \triangleq\left[\begin{array}{ccc}
%\bm{X} & \bm{\epsilon} & \bm{\gamma}\\
%\bm{\epsilon}^{T} & \bm{\eta} & %\bm{\delta}\\
%\bm{\gamma}^{T} & \bm{\delta}^{T} & %\bm{B}
%\end{array}\right]=\left[\begin{array}{c}
%x\\
%\hat{x}\\
%\beta
%\end{array}\right]\left[\begin{array}{c}
%x\\
%\hat{x}\\
%\beta
%\end{array}\right]^{T}=uu^{T}\\
%\hat{\bm{\Gamma}}
% & \triangleq\left[\begin{array}{cccc}
%1 & x^{T} & \hat{x}^{T} & \beta^{T}\\
%x & \bm{X} & \bm{\epsilon} & \bm{\gamma}\\
%\hat{x} & \bm{\epsilon}^{T} & \bm{\eta} & \bm{\delta}\\
%\beta & \bm{\gamma}^{T} & \bm{\delta}^{T} & \bm{B}
%\end{array}\right]=\left[\begin{array}{c}
%1\\
%x\\
%\hat{x}\\
%\beta
%\end{array}\right]\left[\begin{array}{c}
%1\\
%x\\
%\hat{x}\\
%\beta
%\end{array}\right]^{T}.\label{eq: hat_Gamma}
%\end{align}
%\end{comment}
\begin{align}\label{eq: Gamma}
  \hat{\bm{\Gamma}}
 \triangleq\left[\!\!\!
  \begin{blockarray}{cccc}
    1 & x^{T} & \hat{x}^{T} & \beta^{T}\\
    \begin{block}{c[ccc]} 
    x&\bm{X} & \bm{\epsilon} & \bm{\gamma}\\
    \hat{x}&\bm{\epsilon}^{T} & \bm{\eta} & \bm{\delta}\\
    \beta&\undermat{\quad\quad\quad\;\;\,{\bm \Gamma}=uu^T}{\bm{\gamma}^{T} & \bm{\delta}^{T} & \bm{B}}\\ 
    \end{block}
  \end{blockarray}
  \!\right]=\left[\!\!\begin{array}{c}
1\\
x\\
\hat{x}\\
\beta
\end{array}\!\!\right]\!\!\left[\!\!\begin{array}{c}
1\\
x\\
\hat{x}\\
\beta
\end{array}\!\!\right]^{T}\!\!\!\!.
\end{align}

%\begin{align}\label{eq: Gamma}
%  \hat{\bm{\Gamma}}
% \triangleq\!\left[\!\!\!
%  \begin{blockarray}{cccc}
%    1 & x^{T} & \hat{x}^{T} & \beta^{T}\\
%    \begin{block}{c[ccc]} 
%    x&\bm{X} & \bm{\epsilon} & \bm{\gamma}\\
%    \hat{x}&\bm{\epsilon}^{T} & \bm{\eta} & \bm{\delta}\\
%    \beta&\undermat{\quad\quad\quad\;\;\,{\bm \Gamma}=uu^T}{\bm{\gamma}^{T} & \bm{\delta}^{T} & \bm{B}}\\ 
%    \end{block}
%  \end{blockarray}
%  \!\right]\!=\!\left[\!\!\begin{array}{c}
%1\\
%x\\
%\hat{x}\\
%\beta
%\end{array}\!\!\right]\!\!\!\left[\!\!\begin{array}{c}
%1\\
%x\\
%\hat{x}\\
%\beta
%\end{array}\!\!\right]^{T}\!\!\!\!\!=\!\left[\!\!\begin{%array}{c}
%1\\
%u
%\end{array}\!\!\right]\!\!\left[\!\!\begin{array}{c}
%1\\
%u
%\end{array}\!\!\right]^T\!\!\!.
%\end{align}
Notably, the right-most equality in \eqref{eq: Gamma} only holds when ${\rm rank}(\hat{\bm{\Gamma}})={\rm rank}(\bm{\Gamma})=1$. By replacing $uu^T$ in \eqref{eq: Aub_outer} with $\bm{\Gamma}$, the performance verification SDP in Model \ref{model:SDP} can be tightened. Using $\hat{\bm{\Gamma}}$ to define the verification objective function via
\begin{align}
\tilde{e}_{i}(\hat{\bm{\Gamma}})\triangleq{\rm tr}(\bm{M}_{i}\bm{X})-s_{i}^{T}(\bm{W}^{(n)}\hat{x}^{(n)}+b^{(n)}),
\end{align}
the associated tightened SDP is compactly given by
\begin{subequations}\label{eq: full_ST}
\begin{align}\tilde{e}_{i}^{{\star}}=\max_{\hat{\bm{\Gamma}}\succeq0}\;\; & \tilde{e}_{i}(\hat{\bm{\Gamma}})\label{eq: full_ST_obj}\\
{\rm s.t.}\;\; & \bm{A}u-b\ge0\label{eq: ST_lin}\\
 & \bm{A}{\bm \Gamma}\bm{A}^{T}-\bm{A}ub^{T}-bu^{T}\bm{A}^{T}+bb^{T}\ge\bm{0}\label{eq: ST_quad}\\
 & {\rm diag}(\bm{B})=\beta,\label{eq: bin_equality}
\end{align}
\end{subequations}
where $\hat{\bm{\Gamma}}_{1,1}=1$ is implied by \eqref{eq: Gamma}. Notably, the binary equality constraint \eqref{eq: bin_equality} is left in place, because it can't be explicitly captured by Sherali-Adams cuts.
%(although in practice, it can be satisfied even without explicit enforcement).
%\begin{remark}
While \eqref{eq: ST_quad} is originally derived from \eqref{eq: ST_lin}, they are not necessarily redundant constraints, since \eqref{eq: ST_quad} leverages a potentially high-rank SDP matrix variable ${\bm \Gamma}$.
%\end{remark}
%\begin{remark}
%Notably, the binary equality constraint \eqref{eq: bin_equality} is left in place, because it can't be explicitly captured by Sherali-Adams cuts; in practice, it may be implicitly satisfied even without explicit enforcement.
%\end{remark}

%\begin{remark}
%Further tightening can be achieved through additional equality constraints (i.e., \eqref{eq: diag_delta} and \eqref{eq: diag_eta}) which are not explicitly captured by Sherali-Adams cuts.
%\end{remark}

\subsubsection*{Additional Equality Constraints}
Further tightening can be achieved through additional equality constraints which are not explicitly captured by Sherali-Adams cuts. Here, we define two useful ones. First, since a given ReLU output $\hat{x}_{i}$ is equal to 0 if its binary $\beta_{i}$ is 0 or a positive value if $\beta_{i}=1$, then $\beta_{i}\hat{x}_{i}=\hat{x}_{i}$ always holds. Thus, from \eqref{eq: Gamma}, we have
\begin{align}\label{eq: diag_delta}
{\rm diag}(\bm{\delta})=\hat{x}.
\end{align}
Second, from \eqref{eq: quad_equality} in Appendix \ref{AppA}, $\sigma(z)^{2}=z\cdot\sigma(z)$ holds for any ReLU input $z$. Therefore, for any NN layer, where $x$ is an input and $\hat x$ is an output,
\begin{subequations}
\begin{align}
\hat{x}_{i}\hat{x}_{i} & =(\bm{W}_i^Tx+b_i)\hat{x}_{i}\\
 & =\bm{W}_i^T\hat{x}_{i}x+\hat{x}_{i}b_i.
\end{align}
\end{subequations}
Since $\hat{x}_{i}\hat{x}_{i}$ corresponds to diagonal entries of ${\bm \eta}$, and since $\hat{x}_{i}x$ corresponds to a column of ${\bm \epsilon}$, we conveniently have\footnote{To avoid notational complexity, we assume (\ref{eq: diag_eta}) is applied for every NN layer. After the first layer, $\bm{\epsilon}$ must be replaced by properly indexed values from matrix $\bm{\eta}$, since the input to NN layer $i+1$ is actually ${\hat x}^{(i)}$, $i>1$, per \eqref{eq: NN_layers}, and not $x$, which is the very first input to the NN at layer $0$.}
\begin{align}\label{eq: diag_eta}
\bm{\eta}_{i,i}=\bm{W}_{i}^{T}\bm{\epsilon}_{i}+\hat{x}_{i}b_i.
\end{align}

\subsection{Constraint removal \& sequential targeted tightening (STT)}
When combined with \eqref{eq: diag_delta} and \eqref{eq: diag_eta}, formulation \eqref{eq: full_ST} tends to be a very tight relaxation. Unfortunately, since $\bm{A}\in {\mathbb R}^{N \times M}$, \eqref{eq: ST_quad} represents a total of $N^2$ constraint inequalities. For even a moderately small NN (100 neurons), this already becomes an intractable number of constraints (10's of thousands), causing the SDP solver to progress intractably slowly. Accordingly, we propose the utilization of a sequential targeted tightening (STT) approach to handle this problem; similar types of approaches have been very recently applied to the NN verification problem in the ML community~\cite{Ma:2020,Lan:2022}. However, due to the nature of the underlying problem we are trying to solve, our approach is unique in that it exploits iterative tightening to not only query the NN, but to also tighten the SDP matrix variable $\bm X$ associated with querying the ground truth expression the NN is attempting to emulate. That is, ${\rm tr}(\bm{M}_{i}\bm{X})$ in the objective of \eqref{eq: full_ST_obj} outputs the ground truth of the underlying constraint only when ${\rm rank}(\bm{X})=1$. Thus, tightening achieves multiple simultaneous objectives, and it allows us to deliver a method that both maps the NN output \textit{and} determines worst-case deviation from the ground truth. This dynamic arises naturally, in that we are modeling quadratic constraints \eqref{eq: quad_const} with piecewise linear surrogates \eqref{eq: NN_ReLU}.

To design our STT approach, we define the matrix function $\bm{\Omega}:(\mathbb{R}^{t\times t},\mathbb{R}^{t})\rightarrow\mathbb{R}^{N\times N}$ which encodes the LHS of \eqref{eq: ST_quad}:
\begin{align}
{\bm \Omega}(\bm{\Gamma},u)\triangleq\bm{A}\bm{\Gamma}\bm{A}^{T}-\bm{A}ub^{T}-bu^{T}\bm{A}^{T}+bb^{T}.
\end{align}
Set $\mathcal S$ contains the set of all indices associated with matrix $\bm \Omega$:
\begin{align}
\mathcal{S}=\{i,j\,|\,i\in\{1,\ldots,N\},j\in\{1,\ldots,N\}\}.
\end{align}

\begin{lemma}\label{lem_Omega}
If $\bm{A}u-b\ge0$ and ${\rm rank}(\hat{\bm{\Gamma}})=1$, $\hat{\bm{\Gamma}}_{1,1}=1$, then
\begin{align}\label{eq: Omega_ge0}
\bm{\Omega}_{i,j}\ge0,\;\forall \,\{i,j\}\in {\mathcal S}.
\end{align}
\begin{proof}
If $\hat{\bm{\Gamma}}$, $\hat{\bm{\Gamma}}_{1,1}=1$, is a rank one matrix and $\hat{\bm{\Gamma}}_{2:t,1}=u$ (by definition), then by the rank-1 decomposition theorem, $\hat{\bm{\Gamma}}_{2:t,2:t}={\bm{\Gamma}}=uu^T$. Since all elements of $\bm{A}u-b$ are non-negative, and the product of any two non-negative numbers is itself non-negative, then by \eqref{eq: Aub_outer}, relation \eqref{eq: Omega_ge0} must hold.
\end{proof}
\end{lemma}
Unfortunately, the converse of Lemma \ref{lem_Omega} is not true: satisfaction of \eqref{eq: Omega_ge0} does not enforce ${\rm rank}(\hat{\bm{\Gamma}})=1$. If it did, there would exist a polynomial time (SDP) solution to a (nonconvex) NP-hard problem, which generally requires exponential time. However, enforcement of $\bm{\Omega}_{i,j}\ge0$ for any subset of $\mathcal S$ helps limit the relaxed feasible space, thus enabling a tighter relaxed bound $\tilde{e}_{i}^{{\star}}$ on the true performance bound ${e}_{i}^{{\star}}$. Our sequential tightening method hinges on the following observation, which was empirically observed while testing \eqref{eq: full_ST}:
\begin{observation}\label{Obs1}
The enforcement of a small critical subset ${\mathcal S}_c \subset {\mathcal S}$, where $|{\mathcal S}_c| \ll |{\mathcal S}|$, of Sherali-Adams constraints $\bm{\Omega}_{i,j}\ge0$ can result in a fairly tight SDP solution.
\end{observation}
While this observation does not always hold, it serves as the basis of our sequential tightening method. To state this method explicitly, we define the following assessment model, which has resulted from ($i$) relaxing an MIQP to an SDP, ($ii$) tightening the relation by adding exhaustive Sherali-Adams cuts, and ($iii$) loosening the relaxation by keeping only a selective subset of these cuts (the ones in ${\mathcal S}_c$).

\begin{model}[h]
\caption{\hspace{-0.1cm}\textbf{:} NN Performance Assessment via SDP with Tightening from Selective Sherali-Adams Cuts}
\label{model:SA}
\vspace{-0.25cm}
\begin{align}
\tilde{e}_{i}^{{\star}}=\max_{\hat{\bm{\Gamma}}\succeq0}\;\; & \tilde{e}_{i}(\hat{\bm{\Gamma}})\nonumber\\
{\rm s.t.}\;\; & \bm{A}u-b\ge0\tag{Linear Bounds}\label{eq: Scb}\\
 & \bm{\Omega}_{i,j}\ge0,\;\forall \{i,j\}\in\mathcal{S}_{c}\label{eq: Scc}\tag{Sherali-Adams Cuts}\\
 & \eqref{eq: bin_equality},\eqref{eq: diag_delta},\eqref{eq: diag_eta}\tag{Equality Tightening}\label{eq: Scd}.
\end{align}
 \vspace{-0.5cm}
\end{model}

\begin{definition}[{\textbf{Sequential Targeted Tightening (STT)}}] STT solves Model \ref{model:SA} using a small subset $\mathcal{S}_{c}$. It then assesses the worst-case constraint violation across $\mathcal{S}\setminus \mathcal{S}_{c}$ and adds indices associated with the worst constraint violations into $\mathcal{S}_{c}$. Model \ref{model:SA} is then re-solved, and the procedure sequentially iterates.
\end{definition}
The procedure is outlined in 
Alg.~\ref{algo:iterative} using two functions:
\begin{itemize}
    \item ${\rm indmin}({\bm 
    \Omega}<0,N_c)$ returns the indices associated with the $N_c$ smallest values in ${\bm \Omega}$ which are less than 0;
    \item $|\lambda_{1}/\lambda_{2}(\hat{\bm{\Gamma}})|$ returns the absolute ratio\footnote{As ${\rm rank}({\hat{\bm{\Gamma}})}\rightarrow 1$, then $r=|\lambda_{1}/\lambda_{2}(\bm{M})|\rightarrow \infty$. Thus, $r$ serves as an effective measure of matrix $\hat{\bm{\Gamma}}$'s proximity to being rank 1.} of the first and second largest eigenvalues of square matrix $\hat{\bm{\Gamma}}$.
\end{itemize}
Additionally, the active constraint set is initialized on line 1 using two sets: $\mathcal{S}_{\rm dual-feas}$, which is the minimum number of Sherali-Adams constraints needed for preventing dual infeasibility (unbounded objective), and $\mathcal{S}_{\rm diag}$ (defined later), which constrains the diagonal values of the semidefinite matrix $\hat{\bm \Gamma}$.

\begin{algorithm}
\caption{Sequential Targeted Tightening (STT)}\label{algo:iterative}

{\small \textbf{Require:}
Number of constraints $N_c$ to add at each iteration, desired performance threshold $\bar{e}_{i}$, eigenvalue ratio stopping criteria $R_{e}$

\begin{algorithmic}[1]

\State Initialize: $\mathcal{S}_{c}\leftarrow\mathcal{S}_{\rm dual-feas} \cup \mathcal{S}_{\rm diag}$

\State $(\tilde{e}_{i}^{{\star}},\hat{\bm{\Gamma}}^{{\star}},{\bm{\Gamma}}^{{\star}},u^{{\star}})\,\leftarrow$ \textbf{Solve} Model \ref{model:SA} with $\mathcal{S}_{c}$

\While{$\tilde{e}_{i}^{{\star}}\!>\!\bar{e}_{i}$}

\If{$|\lambda_{1}/\lambda_{2}(\hat{\bm{\Gamma}}^{{\star}})|<R_{e}$}

%$|{\mathcal S}|\!>\!|{\mathcal S}_c|$

\State ${\bm \Omega}^{\star}\,\leftarrow$ ${\bm \Omega}(\bm{\Gamma}^{\star},u^{\star})$

\State $\mathcal{S}_{\rm new}\leftarrow$ ${\rm indmin}({\bm \Omega}^{\star}<0,N_c)$

\State $\mathcal{S}_{c}\leftarrow\mathcal{S}_{c}\cup\mathcal{S}_{\rm new}$

\State $(\tilde{e}_{i}^{{\star}},\hat{\bm{\Gamma}}^{{\star}},{\bm{\Gamma}}^{{\star}},u^{{\star}})\,\leftarrow$ \textbf{Solve} Model \ref{model:SA} with updated $\mathcal{S}_{c}$

\Else
\vspace{-2pt}
\State \textit{Terminate} \vspace{3pt}

\EndIf {\bf end}

\EndWhile {\bf end}

\State \Return Performance guarantee $\tilde{e}_{i}^{{\star}}$ with associated solution ${\bm{\Gamma}}^{{\star}}$

\end{algorithmic}}
\end{algorithm}

\begin{figure}
\includegraphics[width=\columnwidth]{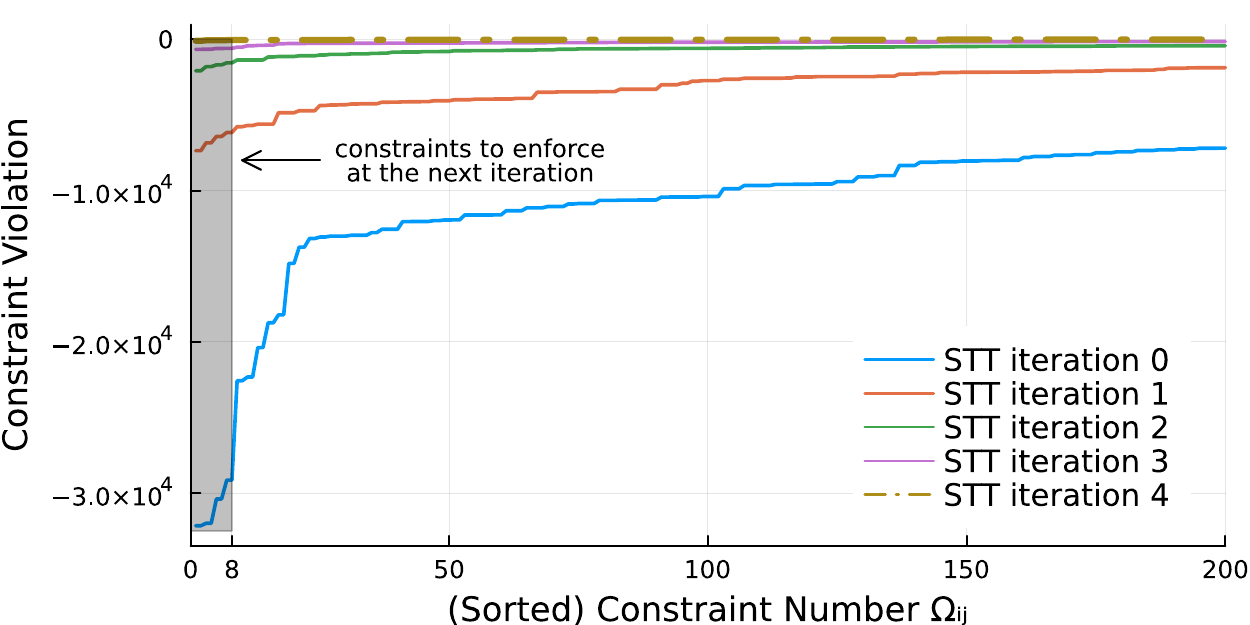}
\caption{Four iterations of the STT algorithm are plotted. For each iteration, the largest 200 (out of 158,404 total) constraint violations are depicted (i.e., where \eqref{eq: Omega_ge0} is violated). At each iteration, the 8 largest constraint violations (gray box) are identified and enforced at the next iteration. After four iterations, the degree of constraint violation becomes minimal.}\label{fig:constraint_violation}
\end{figure}

As stated, Alg.~\ref{algo:iterative} has two general termination criteria:
\begin{enumerate}

\item Performance criteria: if $\tilde{e}_{i}^{{\star}}<\bar{e}_{i}$, the algorithm terminates, since the NN model has been shown to satisfy the minimally acceptable performance criteria $\bar{e}_{i}$.

\item Rank condition: if $|\lambda_{1}(\hat{\bm{\Gamma}}^{{\star}})|/|\lambda_{2}(\hat{\bm{\Gamma}}^{{\star}})|>R_{e}$, the algorithm terminates, since the rank of the SDP solution is sufficiently high and the solution is potentially tight.

% \item Constraint subset: while $|{\mathcal S}|\!>\!|{\mathcal S}_c|$, there are still more constraints that can be added to the active subset ${\mathcal S}_c$.

% \item Constraint satisfaction: if $\mathcal{S}_{\rm new} \ne \emptyset$, no new constraints were added during the previous iteration, so the algorithm should terminate.

\end{enumerate}

Fig.~\ref{fig:constraint_violation} visually depicts four iterations of the STT algorithm\footnote{This example, which will be presented in the test results section, is related to assessing the worst case error of a NN which predicts the current flow in a particular line in the PGLib 14 bus system. There are $398$ linear constraints \eqref{eq: ST_lin} in this system, leading to $398^2$ Sherali-Adams cuts.}. At each iteration, 8 more constraints are added to the subset ${\mathcal S}_c$, causing the solution to iteratively tighten. After just four iterations, 
%the solution eigenvalue ratio $\lambda_1/\lambda_2$ increases by an order of magnitude (from 1.2 to 11.5), and 
the targeted enforcement of just $4\times8 = 32$ extra constraints causes $2,500$ other constraints to be implicitly enforced. This figure highlights the importance of \textit{sequential} constraint enforcement: 
\begin{itemize}
    \item If too many constraints are selected for enforcement at each iteration, many of the selected constraints would have ended up being implicitly enforced anyways, thus unnecessarily slowing down the SDP solver.
    \item However, if too few constraints are enforced, a greater number of SDPs will need to be solved to reach a tight solution, thus slowing down the overall STT procedure.
\end{itemize}
There exists an optimal yet unknown number of constraints which should be selected for enforcement. In this paper, we select this number of constraints based on trial and error.

Each iteration $j$ of Alg.~\ref{algo:iterative} generates a new performance guarantee $\tilde{e}_{i,[j]}^{{\star}}$ which ($i$) bounds the solution of the original non-convex MIQP in Model \ref{model:MIQP} and ($ii$) decreases monotonically in value from the previous iteration's solution $\tilde{e}_{i,[j-1]}^{{\star}}$. This is easily proved in the following theorem.

\begin{theorem}
The performance guarantees generated by each iteration $j> 0$ of Alg.~\ref{algo:iterative} satisfy
\begin{align}\label{eq: bound}
\tilde{e}_{i,[j-1]}^{{\star}}\,\ge\,\tilde{e}_{i,[j]}^{{\star}}\,\ge\,e_{i}^{{\star}}.
\end{align}
\begin{proof}
Denote the feasible solution space of the MIQP (Model \ref{model:MIQP}) as the set ${\mathcal W}$. Further denote the feasible solution spaces of Model \ref{model:SA} in line 8 of Alg.~\ref{algo:iterative}
as $\tilde{\mathcal W}_{[j-1]}$ and $\tilde{\mathcal W}_{[j]}$, associated respectively with iterations $j-1$ and $j$ of the algorithm. Since the constraints in Model \ref{model:SA}
%\eqref{eq: Scb}-\eqref{eq: Scd} 
are relaxed counterparts of the original problem constraints in Model \ref{model:MIQP},
%\eqref{eq: MIQPb}-\eqref{eq: MIQPe},
$\tilde{\mathcal{W}}_{[j]}\supset\mathcal{W},\forall j$. Furthermore, since each iteration of Alg.~\ref{algo:iterative} adds up to $N_c$ constraints to ${\mathcal S}_c$, $\tilde{\mathcal{W}}_{[j-1]}\supset\tilde{\mathcal{W}}_{[j]}$. Since the relaxed feasible space monotonically shrinks around the original feasible space ${\mathcal W}$ (but never precludes any of it), and the objective function does not change between iterations, \eqref{eq: bound} directly follows.
\end{proof}
\end{theorem}

\textit{Targeted constraint enforcement:} Directly enforcing constraints which show the largest violations (i.e., the ones in the gray box in Fig.~\ref{fig:constraint_violation}) is an effective strategy. However, we have observed that more targeted constraint enforcement can help the STT algorithm converge more quickly. In particular, we have observed two key Sherali-Adams cuts:
\begin{enumerate} 
\item The element-wise product of \eqref{eq: ReLU_MILPb} and \eqref{eq: ReLU_MILPd}, yielding $\bm{M}_{i}^{{\rm max}}\beta_{i}{\hat x}_{i}-{\hat x}_{i}^{2}\ge0,\,\forall i$. This implicitly bounds the diagonal values of $\bm \eta$ from \eqref{eq: Gamma}.

\item The element-wise product of the input bounds (which will be introduced in the next section) $v\le{\rm V}^{{\rm {\rm max}}}$ and $-{\rm V}^{{\rm {\rm max}}}\le v$ from \eqref{eq: bound_input}, yielding ${\rm V}^{2,{\rm {\rm max}}} \ge v^{2}$. This implicitly bounds the diagonal values of $\bm X$ from \eqref{eq: Gamma}.

\end{enumerate} 

These two constraints are aggregated into the set ${\mathcal S}_{\rm diag}$ and are used to help initialize the STT algorithm on line 1.

\section{Application to the Power Flow Problem}\label{sec:pf_application}
In this section, we define a learned, piecewise linear NN model which mimics the power flow equations, and then we pose an associated verification problem which STT can solve.
\subsection{Piecewise linear power flow model}
To pose this model, we consider a power system network whose graph ${\mathcal G}({\mathcal V},{\mathcal E})$ has edge set $\mathcal{E}$, $|\mathcal{E}|=n_l$, vertex set $\mathcal{V}$, $|\mathcal{V}|=n_b$, and signed nodal incidence matrix $E\in{\mathbb R}^{n_l\times n_b}$. The complex nodal admittance matrix, known as the Y-bus matrix, is ${Y}_b\in{\mathbb C}^{n_b\times n_b}$. Our learned surrogate model ${\hat f}(\cdot)$ maps network nodal voltages in rectangular coordinates ($v_r,v_i$) to power flow quantities of interest: voltage magnitude squared (${\rm V}^{2}$), active and reactive power injection ($p^{{\rm inj}}$, $q^{{\rm inj}}$), and current flow squared on both directions of each line in the network ($l^{{\rm ft}}$, $l^{{\rm tf}}$). Thus,
\begin{align}\label{eq: nn_map}
\hat{f}(\underbrace{v_{r},v_{i}}_{x})\rightarrow(\underbrace{{\rm V}^{2},p^{{\rm inj}},q^{{\rm inj}},l^{{\rm ft}},l^{{\rm tf}}}_{y}).
\end{align}
The imaginary voltage at the reference bus is set to 0 and excluded from the voltage vector. Therefore, the NN maps input $x\in {\mathbb R}^{2n_b-1}$ to output $y\in {\mathbb R}^{3n_b+2n_l}$. Similar algebraic surrogate mappings have been recently proposed and verified for the ACOPF problem~\cite{eydenberg4245628physics}. Explanations of how this model can be used to verify, e..g, NN ACOPF solvers are presented in the conclusion section. We focus on this particular mapping for three reasons: ($i$) it captures the power flow quantities which are typically needed to solve OPF and UC type problems, ($ii$) it directly mimics the forward mappings of QCQP (e.g., $p=v^TM_pv$) and SDP (e.g., $p={\rm tr}(M_pvv^T)$) OPF and UC formulations, and ($iii$) other works which learn or verify over this sort of mapping~\cite{eydenberg4245628physics, Kody:2021}.

\subsection{Power flow model verification}
To formulate the verification problem, we denote the constant matrices $\bm{M}_{{\rm V},i}$, $\bm{M}_{p,i}$, $\bm{M}_{q,i}$, $\bm{M}_{f,i}$, and $\bm{M}_{t,i}$ which map the input vector $x$ to elements of the output vector $y$ in \eqref{eq: nn_map} via, e.g., ${\rm V}_{i}^{2}={\rm tr}\{\bm{M}_{{\rm V},i}xx^{T}\}$, $p_{i}^{{\rm inj}}={\rm tr}\{\bm{M}_{p,i}xx^{T}\}$, etc. These matrices are typically used in the QCQP and SDP OPF formulations~\cite{Wu:2018}. Assuming the NN has been reformulated and suitable big-M bounds have been identified, input and output bounds are the final elements needed to construct Model \ref{model:SA}. These bounds depend on the region over which we would like to verify the model's performance (e.g., OPF):
\begin{subequations}\label{eq: bound_output}
\begin{align}
{\rm V}^{2,{\rm {\rm min}}} & \le{\rm V}^{2}\le{\rm V}^{2,{\rm max}}\label{eq: v_lim}\\
p_{G}^{{\rm min}}-p_{D}^{{\rm max}} & \le p^{{\rm inj}}\le p_{G}^{{\rm max}}-p_{D}^{{\rm min}}\label{eq: p_lim}\\
q_{G}^{{\rm min}}-q_{D}^{{\rm max}} & \le q^{{\rm inj}}\le q_{G}^{{\rm max}}-q_{D}^{{\rm min}}\label{eq: q_lim}\\
l^{{\rm ft},{\rm min}} & \le l^{{\rm ft}}\le l^{{\rm ft},{\rm max}}\\
l^{{\rm tf},{\rm min}} & \le l^{{\rm tf}}\le l^{{\rm tf},{\rm max}}\label{eq: ltf_lim}.
\end{align}
\end{subequations}
Bounds on the input may be stated similarly:
\begin{subequations}\label{eq: bound_input}
\begin{align}
-{\rm V}^{{\rm {\rm max}}} & \le v_{r}\le{\rm V}^{{\rm {\rm max}}}\\
-{\rm V}^{{\rm {\rm max}}} & \le v_{i}\le{\rm V}^{{\rm {\rm max}}}.
\end{align}
\end{subequations}
%Stronger bounds on the inputs may be identified using, e.g., the bound tightening methods in~\cite{Coffrin:2017}, but they are not considered in this paper. 
Using these bounds, the NN reformulation (Appendix \ref{AppA}), and \eqref{eq: Aub1}-\eqref{eq: Aub4}, Model \ref{model:SA} may be fully constructed for a given verification index $i$, and the STT procedure may be applied. 

\section{Testing Setup}\label{sec:test_setup}
This section sets up the numerical tests.
%Accordingly, this section builds a data collection procedure for generating NN training data, and it defines the NN structure and training procedure. 
%All code developed in this paper is posted on GitHub~\cite{Chevalier_github:2024}.

\subsection{Training data collection}
Supervised ML algorithms need large and balanced datasets in order to produce high quality models. To generate this data, we utilize an optimization-based approach developed in~\cite{Ventura:2022}, where a nonlinear objective function is maximized in order to find successive power flow solutions which are maximally far apart. To do this, we define sets which are the union of loads and generators: $\mathcal{P}=p_{G}\cup p_{D}$ and $\mathcal{Q}=q_{G}\cup q_{D}$. Next, we define a dataset $\mathcal D$ which initially contains a single load flow solution, given as ${\mathcal D}=(P^{0},Q^{0},{\rm V}^{0})$. The objective function $f_d(p,q,{\rm v})$ quantifies the distance between the given point $(p,q,{\rm v})$ and previous points within the dataset ${\mathcal D}$ via
\begin{align}
f_d(p,q,{\rm v})= & \sum_{j\in\mathcal{D}}\Big(\log\sum_{i\in\mathcal{P}}|p_{i}-P_{i}^{(j)}|+\log\sum_{i\in\mathcal{Q}}|q_{i}-Q_{i}^{(j)}|\nonumber\\
 & \quad\quad\quad\;+\log\sum_{i\in{\mathcal{V}}}|{\rm v}_{i}-{\rm V}_{i}^{(j)}|\Big).
\end{align}
We then maximize this function subject to OPF limits:
\begin{subequations}\label{eq: data}
\begin{align}\max_{p,q,{\rm v}}\;\; & f_d(p,q,{\rm v})\\
{\rm s.t.}\;\;
& p^{{\rm inj}}+q^{{\rm inj}}=v\odot\left(Y_{b}v\right)^{*}\\
& 
l^{{\rm ft/tf}}=\left(Y_{{\rm ft/tf}}v\right)\odot\left(Y_{{\rm ft/tf}}v\right)^{*}\\
& \text{system limits: } \eqref{eq: v_lim}-\eqref{eq: ltf_lim}.
%\text{power flow physics}
\end{align}
\end{subequations}
Formulation \eqref{eq: data} differs from a conventional OPF in two key ways: first, the cost function optimizes for distance from other solutions, as opposed to generation price, and second, the system limits allow for load set-points to vary, according $p_D^{\rm min/max}$ and $q_D^{\rm min/max}$ in \eqref{eq: p_lim}-\eqref{eq: q_lim}. Each time \eqref{eq: data} is solved, the solution is added to the dataset ${\mathcal D}$. This data is then used for training the NN.

\subsubsection*{Practical considerations} 
\begin{figure}
\includegraphics[width=1.0\columnwidth]{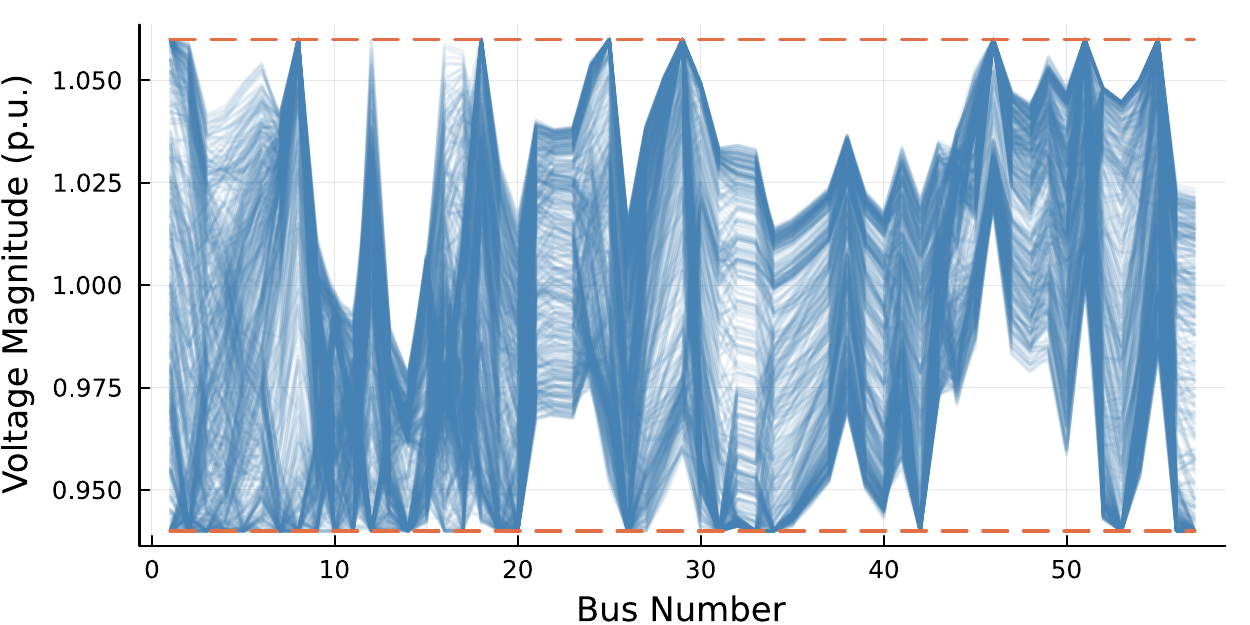}
\caption{500 voltage magnitude solutions are plotted for the 57 bus PGLib system, where each solution is associated with a single iteration of Alg. \ref{algo:data_collect} and a corresponding maximization of \eqref{eq: data}. The upper and lower  voltage profile outlines form an approximate epigraph of the feasible voltage space.}\label{fig:voltage_profile}
\end{figure}

Formulation \eqref{eq: data} can be solved with the interior point solver IPOPT very efficiently. For systems with many hundreds of buses, solving \eqref{eq: data} can become very slow as the dataset $\mathcal D$ grows in size. Accordingly, to accelerate the program, we make two modifications. First, we use a sampled subset of ${\mathcal D}_s \subset {\mathcal D}$ to include in the objective function, and second, we only measure the distances across a sampled subset of variables: ${\mathcal P}_s\subset{\mathcal P}$, ${\mathcal Q}_s\subset{\mathcal Q}$, and ${\mathcal V}_s\subset{\mathcal V}$. Using these updates, the formal data collection procedure is outlined in Alg. \ref{algo:data_collect}, which is solved using a custom modification of PowerModels.jl~\cite{Coffrin:2018} and the IPOPT solver; please refer to our code for details~\cite{Chevalier_github:2024}. The voltage magnitude solutions associated with 500 iterations of Alg.~\ref{algo:data_collect} (i.e., 500 feasible OPF solutions) for the 57-bus PGLib~\cite{Babaeinejadsarookolaee:2019} system are depicted in Fig.~\ref{fig:voltage_profile}. In this experiment, all loads were allowed to vary by $\pm 50\%$ from their PGLib nominal values. Table \ref{tab:collection} summarizes the test cases considered in the paper and the number of power flow samples we collected from them.

%~\cite{wachter2006implementation}

\begin{table}[h]
   \captionsetup{justification=centering}
   \caption{Data Collection Characteristics} 
   \label{tab:collection}
   \small
   \centering
   \begin{tabular}{c|ccccc}
   \toprule\toprule
  \textbf{PGLib} & {power flow}    & {total inputs}   & {total outputs}   & {load} \vspace{-0.35mm}\\  
  \textbf{Network} & {data samples}    & {$(2n_b\!-\!1)$}   & {$(3n_b\!+\!2n_l)$}   & {range}
  \\
   \midrule 
   14-bus & 250 & 27 & 82 & $\pm 50\%$\\
   57-bus & 500 & 113 & 331 & $\pm 50\%$\\
   118-bus & 750 & 235 & 726 & $\pm 50\%$\\
   200-bus & 1000 & 399 & 1090 & $\pm 50\%$\\
\bottomrule
   \end{tabular}
\end{table}

\begin{algorithm}
\caption{Optimization-Based Data Collection}\label{algo:data_collect}

{\small \textbf{Require:}
Number of data points to collect $N_d$, number of sampled variables ($n_{v}$) and datapoints ($n_{d}$) to include in objective function, initial OPF solution, $k=1$ ${\mathcal D}=(P^{0},Q^{0},{\rm V}^{0})$

\begin{algorithmic}[1]

\While{$k\le N_d$}

\State $({\mathcal P}_s,{\mathcal Q}_s,{\mathcal V}_s)\leftarrow$ sample $n_v$ variables from $({\mathcal P},{\mathcal Q},{\mathcal V})$

%\State Sample $p$ variables: ${\mathcal P}_s$ = sample(${\mathcal P}$,$n_v$)

%\State Sample $q$ variables: ${\mathcal Q}_s$ = sample(${\mathcal Q}$,$n_v$)

%\State Sample $\rm v$ variables: ${\mathcal V}_s$ = sample(${\mathcal V}$,$n_v$)

\State ${\mathcal D}_s\leftarrow$ sample $n_d$ solutions from data library ${\mathcal D}$

\State $(P^{(k)},\!Q^{(k)},\!{\rm V}^{(k)})\leftarrow$ \textbf{solve} \eqref{eq: data} for subsets ${\mathcal P}_s$, ${\mathcal Q}_s$, ${\mathcal V}_s$, ${\mathcal D}_s$

\State update library: $\mathcal{D}=\mathcal{D}\cup(P^{(k)},Q^{(k)},{\rm V}^{(k)})$

\State $k \leftarrow k + 1$

\EndWhile {\bf end}

\State Split $\mathcal{D}$ into train (70\%), validate (15\%), and test (15\%) sets

\State \Return $\mathcal{D}_{\rm train}$, $\mathcal{D}_{\rm val}$, $\mathcal{D}_{\rm test}$

\end{algorithmic}}
\end{algorithm}

\subsection{Neural network architecture and training}
In this paper, we employ a double layer NN, where the first hidden layer is equipped with ReLU activation functions, and output layer is equipped with linear activation functions: $y=\bm{W}^{(1)}\sigma(\bm{W}^{(0)}x+b^{(0)})+b^{(1)}$. Notably, this is the smallest NN which still satisfies the so-called universal approximation theorem~\cite{Leshno:1993}, so its representational power is not limited. To exploit physics, however, and limit the complexity of the NN, we train the NN to learn corrections to a physics-based linearization, as in~\cite{Kody:2021}. This additional affine feedthrough term consists of a Jacobian $\bm{J}^{\star}$ and residual $r^{\star}$:
%Hornik:1989,
\begin{align}
y=\bm{W}^{(1)}\sigma(\bm{W}^{(0)}x+b^{(0)})+b^{(1)}+\bm{J}^{\star}x+r^{\star}.
\end{align}
The explicit construction of the linear inequality $\bm{A}u-b\ge0$, and the inclusion of input and output normalization layers, is given on the GitHub associated with this paper:~\cite{Chevalier_github:2024}
%we also show the inclusion of input and output normalization layers
%The explicit construction of the linear inequality $\bm{A}u-b\ge0$ is given on the GitHub page associated with this paper:~\cite{Chevalier_github:2024}
%In Appendix \ref{AppB}, we also show the inclusion of input and output normalization layers, and we explicitly transform the system of equations into a set of linear inequalities: $\bm{A}u-b\ge0$.

\begin{figure}
\includegraphics[width=\columnwidth]{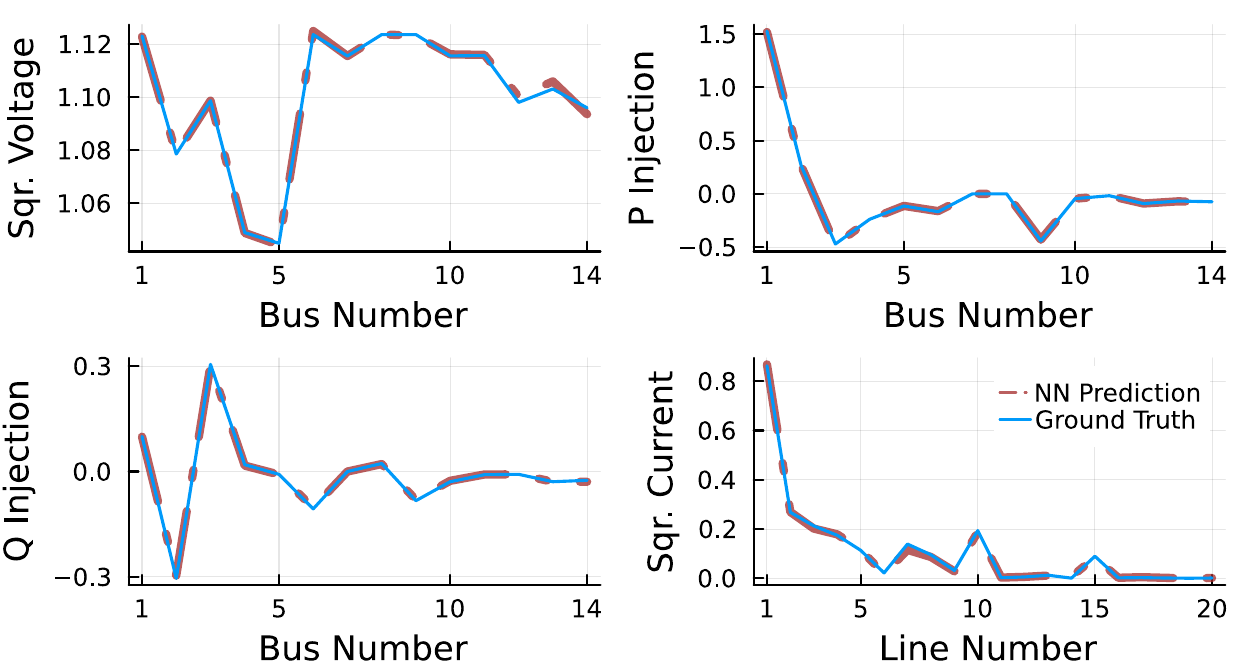}
\caption{The prediction of a trained NN (trained on data from the PGLib 14 bus system) is plotted against the ground truth power flow data for a single test point (i.e., a point not used to train the NN).}
\label{fig:nn}
\end{figure}

All training was performed with Adam~\cite{Kingma:2014} in Flux.jl%~\cite{Innes:2018} 
with randomly shuffled data batching. Training hyperparameters are summarized in Table \ref{tab:training}. Cross-validation (using 15\% of the data withheld for validation) was used to select the best model across all training epochs. After half of the training epochs, weight matrices were ``sparsified" by setting a certain percentage (see Table \ref{tab:training}) of model parameters to 0 (and holding at 0 for the remainder of the training). Finally, at the conclusion of training, all model parameters smaller than the threshold $\epsilon = 0.001$ were set to 0, incurring a drop in validation accuracy of less than 0.1\%. This final step greatly enhanced the numerical stability of the performance assessment algorithm. Fig.~\ref{fig:nn} depicts a trained NN output.

\begin{table}[h]
   \captionsetup{justification=centering}
   \caption{Neural Network Training Parameters} 
   \label{tab:training}
   \small
   \centering
   \begin{tabular}{c|ccccc}
   \toprule\toprule
  \textbf{PGLib} & {ReLU}    & {training}   & {learning}   & {weight} & {batch} \vspace{-0.35mm}\\  
  \textbf{Network} & {number}    & {epochs}   & {rate}   & {sparsity} & {size}
  \\
   \midrule 
   14-bus & 25 & 80k & 2e-4 & 50\% & 15\\
   57-bus & 50 & 80k & 5e-4 & 50\% &  25 \\
   118-bus & 75 & 50k & 10e-4 & 75\% & 50 \\
   200-bus & 100 & 50k & 10e-4 & 75\% & 75 \\
\bottomrule
   \end{tabular}
\end{table}

Computing tight values for big-M parameters in the NN-to-MILP reformulation of \eqref{eq: ReLU_MILPc}-\eqref{eq: ReLU_MILPd} is a computationally challenging but important step. In this paper, we used a three-step approach developed in~\cite{Venzke:2020} to compute these big-M values; see~\cite{Venzke:2020} and \cite[eq.~14]{Kody:2021} for details.

%\subsection{Bounding Big-M}
%Computing tight values for big-M parameters in the NN-to-MILP reformulation of \eqref{eq: ReLU_MILPc}-\eqref{eq: ReLU_MILPd} is a computationally challenging but important step. In this paper, we used a three step approach developed in~\cite{Venzke:2020}, which is summarized below:

%\begin{enumerate}
    %\item First, we used interval arithmetic and the input bounds \eqref{eq: bound_input} to compute the maximum and minimum ReLU inputs; these served as loose big-M estimates.
    
    %\item Using these estimates, we then solved an LP relaxation of the full MILP subject to output and input bounds \eqref{eq: bound_output}-\eqref{eq: bound_input} to get even tighter big-M estimates.
    
    %\item Finally, using these tighter estimates, we solved the un-relaxed MILP to get optimally tight big-M values. See~\cite[eq.~14]{Kody:2021} for an example of this formulation.
%\end{enumerate}

%Solving step 3 to 0\% optimality gap for every ReLU is very time consuming, so we set a time limit of 75 seconds for each big-M computation, still requiring almost 5 hours of computation. All MILPs were solved with Gurobi 11.0.

\section{Numerical Test Results}\label{sec:results}
In this section, we assess the performance of STT by comparing its error bounds (i.e., performance guarantees) with ones generated by the Gurobi 11.0 MIQP solver; we generate these guarantees for the NNs trained in the previous section. All test code is posted on GitHub~\cite{Chevalier_github:2024}.
%Second, we consider the relationship between eigenvalue ratios and solutions tightness. Finally, we provide a discussion concerning the performance and limitations of STT.
%\subsection{Exhaustive model comparisons}
To run these tests, we selected particular NN output predictions from the trained model \eqref{eq: nn_map}, and then we allowed STT Alg.~\ref{algo:iterative} to take 10 iterations, using MOSEK 10.1 to solve the SDPs. At each iteration, we added $N_c=150$, $250$, $300$, and $350$ new Sherali-Adams cuts to the model (for the 14, 57, 118, and 200-bus systems, respectively). After running STT, we allowed Gurobi 11.0's non-convex MIQP solver to attempt solving the equivalent, non-relaxed problem (posed as Model \ref{model:MIQP}) for an equal amount of time. As Gurobi's Branch and Bound (BaB) algorithm runs, it generates two bounds: a lower bound based on the best incumbent candidate solution, and a worst case upper bound. We track this upper bound and compare it to the one generated by STT. All loads could vary by $\pm50\%$. In
the following analysis, we focus on neither the value of the bound nor its quality, since the former is dependent on training and data collection routines, and the latter is subjectively dependent on the intended model application.

\begin{figure}
\includegraphics[width=\columnwidth]{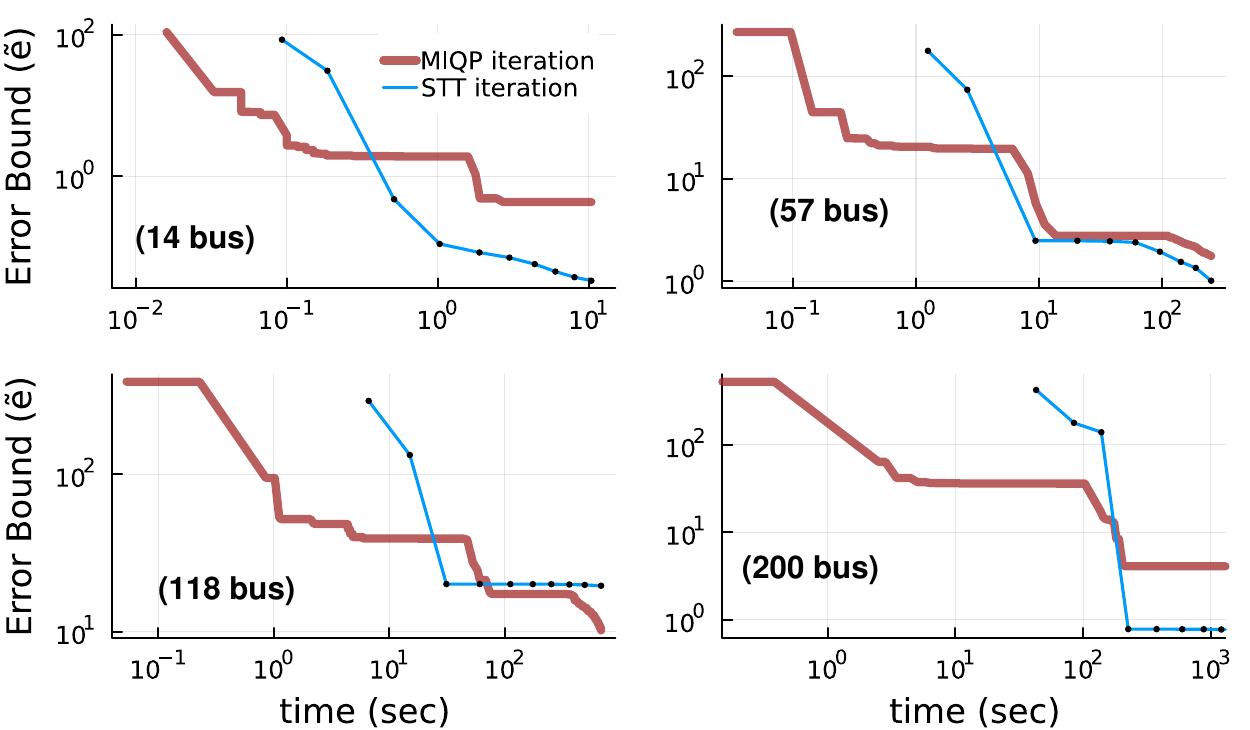}
\caption{Performance guarantees (worst case underestimation of active power injection at load buses) generated by ($i$) Gurobi's MIQP solver (red) through BaB iterations, and ($ii$) 10 iterations of STT algorithm (blue).}
\label{fig:power_bound_race}
\end{figure}

\begin{figure}
\includegraphics[width=\columnwidth]{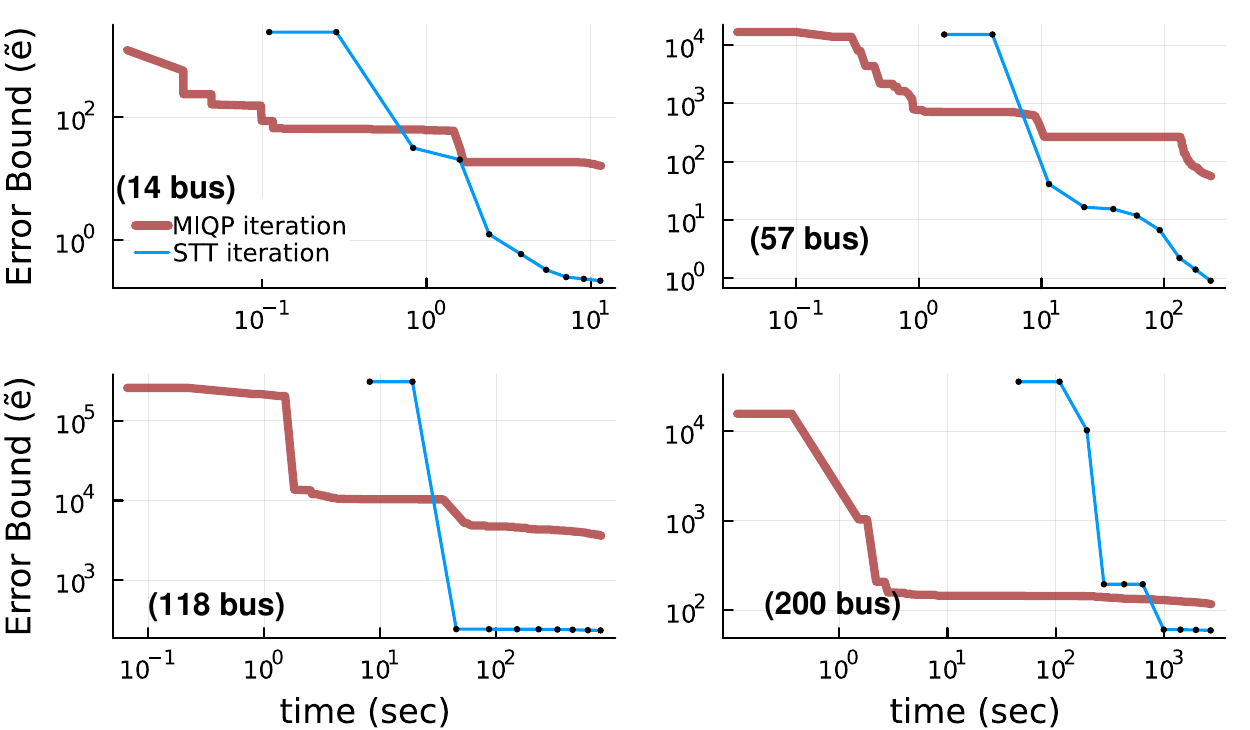}
\caption{Performance guarantees (worst case overestimation of square current flow on the network line with the largest ampacity) generated by ($i$) Gurobi's MIQP solver (red), and ($ii$) 10 iterations of STT algorithm (blue).}
\label{fig:current_bound_race}
\end{figure}

Direct comparison of the STT iterations and Gurobi's upper bound are shown in Fig. \ref{fig:power_bound_race}, which compares active power injection error at load buses. Similarly, Fig. \ref{fig:current_bound_race} compares square current flow error on the line in each network with the largest ampacity. Across each PGLib case in both tests, STT typically overtook the MIQP bound within just several iterations. At the end of the time budgets, the MIQP's worst case model error bound $\tilde{e}_{{\rm MIQP}}$ was anywhere from slightly smaller to orders of magnitude larger than $\tilde{e}_{{\rm STT}}$, the bound generated by STT. Notably, STT iterations slowed down considerably as more and more constraints were added to the SDP. Table \ref{tab:compareSolvers} provides more exhaustive results. In it, we track final bound ratio
\begin{align}
    \tau=\tilde{e}_{{\rm MIQP}}/\tilde{e}_{{\rm STT}}.
\end{align}
In this table, we solved NN verification problems for active and reactive power injections at five buses across each PGLib network; buses were selected as ones which have the highest degree of interconnection with surrounding buses. NN verification problems at these buses tend to be the hardest, since they include the largest number of voltage variable terms in the objective function. Across most trials, STT was able to produce NN verification bounds which were significantly tighter than what MIQP could produce in an equivalent amount of time.

\begin{table}
   \caption{Comparison of MIQP and STT bounds ($\tau=\tilde{e}_{{\rm MIQP}}/\tilde{e}_{{\rm STT}}$)} 
   \label{tab:compareSolvers}
   \small
   \centering
   \begin{tabular}{c|ccccc}
   \toprule\toprule
   \multirow{2}{4em}{$\;\,$\textbf{PGLib \\Network}} & \textbf{error to} & max & min & mean & mean \vspace{-0.35mm}\\  
				       & \textbf{bound} & $\tau$ & $\tau$ & $\tau$ & runtime\\ 

   \midrule 
   \multirow{2}{4em}{$\;$14-bus} & $p_i^{\rm inj}$   & 24.0  & 0.98   & 10.1  & 11.6s  \\
                             & $q_i^{\rm inj}$ & 13.2  & 1.5  & 8.5 & 12.2s \\
   \midrule 
   \multirow{2}{4em}{$\;$57-bus} & $p_i^{\rm inj}$ & 8.0   & 0.29    & 3.7 & 273.7s    \\
                             & $q_i^{\rm inj}$ & 6.5 & 0.37 & 2.1 & 325.4s \\

   \midrule 
   \multirow{2}{4em}{118-bus} & $p_i^{\rm inj}$ & 5.9   & 0.48    & 2.2 & 772.5s   \\
                              & $q_i^{\rm inj}$  & 1.8 & 0.4 & 0.91 & 836.45s\\

   \midrule 
   \multirow{2}{4em}{200-bus} & $p_i^{\rm inj}$ & 1880.0 & 5.4 & 948.6 & 2296.7s    \\
                             & $q_i^{\rm inj}$  & 617.0 & 1.8 & 273.0 & 2559.5s  \\
\bottomrule
   \end{tabular}
\end{table}

\begin{figure}
\includegraphics[width=\columnwidth]{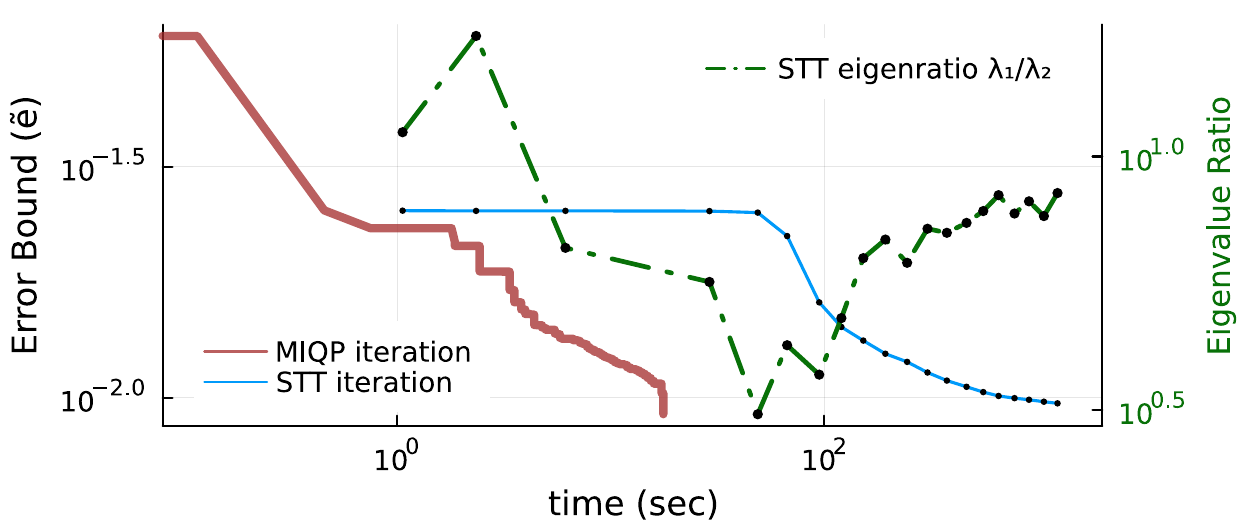}
\caption{Performance guarantees (worst case overestimation of square voltage on bus 1 of the 57 bus network) generated by ($i$) Gurobi's MIQP solver (red), and ($ii$) 20 iterations of the STT algorithm (blue). The eigenvalue ratio $\lambda_1({\hat{\bm \Gamma}})/\lambda_2({\hat{\bm \Gamma}})$ is plotted in green.}
\label{fig:eigen_plot}
\end{figure}

There were a number of situations, however, where MIQP outperformed STT; some of these situations tend to coincide with MIQP objective functions which are naturally convex. For example, the constant matrix ${\bm M}_{{\rm V},j}$ which maps rectangular voltage to square voltage magnitude
%, i.e.,
%\begin{align}
%{\rm V}_{j}^{2}=\left[\!\!\begin{array}{cc}
%v_{r,j} \!\!& v_{i,j}\end{array}\!\!\right]\left[\!\!\begin{array}{cc}
%1 & 0\\
%0 & 1
%\end{array}\right]\left[\!\!\begin{array}{c}
%v_{r,j}\\
%v_{i,j}
%\end{array}\!\!\right]
%\end{align}
has two positive eigenvalues. Thus, the associated overestimation verification problem is convex in its quadratic terms (not in integer variables, though). Fig.~\ref{fig:eigen_plot} shows a race between MIQP and STT. MIQP reached the global solution (with 0\% gap) almost two orders of magnitude faster than STT did.

On the right-side axis in Fig.~\ref{fig:eigen_plot}, we also plot the eigenvalues of matrix $\hat{\bm \Gamma}$ from \eqref{eq: Gamma} as STT iterated. Notably, even when the STT performance bound was within 5\% of the global solution, the SDP matrix was still very far from being rank 1. This suggests that rank 1 solutions might not be necessary for achieving acceptably tight upper bounds on this class of performance verification problem. It also might suggest, however, some form of solution space degeneracy. The authors believe this to be a plausible explanation, since the lifted binary SDP matrix $\bm B$ could potentially satisfy the NN mapping equations, equality constraint \eqref{eq: bin_equality}, and all Sherali-Adams cuts without being rank 1. Future work will explore this question.

\section{Conclusion}\label{sec:conclusion}
Computing worst-case violations for NNs which model the AC power flow equations is a hard problem. In this paper, we have proposed an approach, termed Sequential Targeted Tightening (STT), which solves a semidefinite relaxation of this problem. In order to tighten the solutions computed by the SDP solver, we generate an exhaustive number of Sherali-Adams cuts, and we sequentially enforce the cuts which show the highest degree of violation after each SDP solve. STT can generate worst-case violation bounds which are often over an order of magnitude tighter than the bounds generated by a state-of-the-art MIQP solver (for a given time budget). In power systems, where performance guarantees are essential for deploying NNs to real applications, STT shows great promise.

\subsubsection*{Model extensions} While this paper focused on verifying the surrogate mapping \eqref{eq: nn_map}, STT can be applied to a host of other power flow verification problems. We demonstrate by considering a standard learned OPF mapping~\cite{NELLIKKATH2022108412}: ${\rm NN}(p_d, q_d)\rightarrow (p_g, {\rm V}_g)$. To ensure that a predicted set-point never violates, e.g., a line flow limit, we may pose the following MIQCP:
\begin{subequations}
\begin{align}
\max_{p_d,q_d}\;\, & |l_{i}^{{\rm ft}}-\overline{l}_{i}^{{\rm ft}}|\\
{\rm s.t.}\;\, & {\rm NN}(p_{d},q_{d})\rightarrow(p_{g},{\rm V}_{g})\\
 & v^{T}M_{\{p,q,l,{\rm V}\}}v=\{p,q,l,{\rm V}\}.
\end{align}
\end{subequations}
By ($i$) convexifying the QC power flow constraints via ${\rm tr}(Mvv^{T})$, ($ii$) transforming the NN mapping via binary relaxation, and ($iii$) lifting the power injection variables, STT can be directly applied to this challenging verification problem.

\subsubsection*{Model limitations} While STT showed promising performance, we note three salient challenges for future work:

\begin{itemize}
    \item \textit{Numerical sensitivity}: the greatest challenge associated with running Alg.~\ref{algo:iterative} was numerical sensitivity (ill conditioning). NN model parameters range over orders of magnitude, and generating Sherali-Adams cuts accentuates this problem. To prevent MOSEK from stalling, we increased relevant numerical tolerances to $10^{-6}$ and still ran into convergence problem in several tests.
   
    \item \textit{Scalability}: The largest SDP matrix solved for in this paper was $599\times 599$. Solving significantly larger SDPs will be prohibitive, so alternative SDP relaxation techniques (e.g., the Lagrangian SDP relaxation in~\cite{Dvijotham:2020}) and sparsity exploiting methods will need to be explored.
    
    \item \textit{Smarter constraint enforcement}: the speed and success of Alg.~\ref{algo:iterative} was greatly dependent upon the number of constraints $N_c$ enforced at each iteration. Methods for determining which, and how many, additional constraints to enforce at each iteration are key.
    
\end{itemize}

\appendices

{\section{}\label{AppA}}
We review two methods commonly used to reformulate ReLU-based NNs into equivalent optimization constraints.

\subsection{Mixed Integer Linear Program (MILP) reformulation}
The ReLU-based NN can be transformed into an equivalent MILP~\cite{Venzke:2020}. For example, a single layer NN can be recast as
\begin{subnumcases}
{{\hat x}=\sigma(\bm{W}x+b)\;\Leftrightarrow\;}
{\hat x} \ge\bm{W}x+b\label{eq: ReLU_MILPa}\\ 
{\hat x} \ge0\label{eq: ReLU_MILPb}\\
{\hat x} \le\bm{W}x+b-\bm{M}^{{\rm min}}({\bf 1}-\beta)\label{eq: ReLU_MILPc}\\
{\hat x} \le\bm{M}^{{\rm max}}\beta\label{eq: ReLU_MILPd}\\
\beta \in\{0,1\}^{b},
\end{subnumcases}
where $\beta$ is a vector of binary variables, ${\bf 1}$ is a vector of all ones, and $\bm{M}^{{\rm min}}$, $\bm{M}^{{\rm max}}$ are diagonal matrices filled with ``big-M" constants which bound, respectively, the largest possible output and the smallest possible input to the ReLU operator. By rearranging \eqref{eq: ReLU_MILPa}-\eqref{eq: ReLU_MILPd}, we may restate the constraints associated with the 1-layer ($1l$) NN more compactly via 
\begin{align}\label{eq: relu_to_MILP}
\bm{A}_{{\rm nn}}^{(1l)}\left[\begin{array}{c}
x\\
{\hat x}\\
\beta
\end{array}\right]\ge b_{{\rm nn}}^{(1l)},
\end{align}
where $\bm{A}^{(1l)}_{{\rm nn}}$ and $b^{(1l)}_{{\rm nn}}$ are given by
\begin{align*}
\bm{A}^{(1l)}_{{\rm nn}}=\left[\!\!\begin{array}{ccc}
-\bm{W} & \bm{I} & \bm{0}\\
\bm{0} & \bm{I} & \bm{0}\\
\bm{W} & -\bm{I} & \bm{M}^{{\rm min}}\\
\bm{0} & -\bm{I} & \bm{M}^{{\rm max}}
\end{array}\!\!\right],\,b^{(1l)}_{{\rm nn}}=\left[\!\!\begin{array}{c}
b\\
0\\
\bm{M}^{{\rm min}}{\bf 1}-b\\
0
\end{array}\!\!\right].
\end{align*}
We note that (\ref{eq: relu_to_MILP}) may be extended to accommodate any arbitrary number of NN layers. This is accomplished by defining a series of transformations
\begin{subequations}\label{eq: NN_layers}
\begin{align}
\hat{x}^{(1)} & =\sigma(\bm{W}^{(0)}x+b^{(0)})\\
\hat{x}^{(2)} & =\sigma(\bm{W}^{(1)}\hat{x}^{(1)}+b^{(1)})\\[-6pt]
 & \;\;\vdots\nonumber \\[-5pt]
\hat{x}^{(n)} & =\sigma(\bm{W}^{(n-1)}\hat{x}^{(n-1)}+b^{(n-1)}).
\end{align}
\end{subequations}
A series of single layer NN matrices and vectors can then be combined together into $\bm{A}_{{\rm nn}}$ and $b_{{\rm nn}}$ which collectively obey
\begin{align}\label{eq: Aub}
    \bm{A}_{{\rm nn}}u\ge b_{{\rm nn}}.
\end{align}
In \eqref{eq: Aub}, $u$ contains the NN input $x$, the intermediate NN variables $\hat x$, and all of the binary variables $\beta$:
\begin{align}
u^{T}=[x^{T}\,\hat{x}_{1}^{T}\,\cdots\,\hat{x}_{n}^{T}\,\beta^{T}].
\end{align}
The transformation from \eqref{eq: NN_ReLU} is completed via $y=\bm{W}^{(n)}\hat{x}^{(n)}+b^{(n)}$. To ``select" the $i^{\rm th}$ output of $y$, such that $y_i=s_{i}^Ty$, we define section vector $s_i \in {\mathbb R}^m$ whose $j^{\rm th}$ value is defined by $s_{i_{j}}=0,\;j\ne i$ and $s_{i_{j}}=1,\;j=i$.
%\begin{align}\label{eq: selection}
%s_{i_{j}}=\begin{cases}
%0, & j\ne i\\
%1, & j=i.
%\end{cases}
%\end{align}

\subsection{Quadratic reformulation}
The ReLU-based NN can also be transformed into an equivalent program with quadratic and linear constraints~\cite{Dvijotham:2020}:
\begin{subnumcases}
{{\hat x}=\sigma(\bm{W}x+b)\;\Leftrightarrow\;}
{\hat x}\odot {\hat x} ={\hat x}\odot\left(\bm{W}x+b\right)\label{eq: quad_equality}\\
{\hat x} \ge \bm{W}x+b\\
{\hat x} \ge0,
\end{subnumcases}
where $\odot$ is the element-wise Hadamard product. In~\cite{Dvijotham:2020}, this formulation is tightened by adding the constraint $\left(\hat{x}-\mu_{\hat{x}}\right)\odot\left(\hat{x}-\mu_{\hat{x}}\right)\le\epsilon_{\hat{x}}\odot\epsilon_{\hat{x}}$, where $\mu_{{\hat x}}=\tfrac{1}{2}(l_{{\hat x}}+u_{{\hat x}})$, $\epsilon_{{\hat x}}=\tfrac{1}{2}(l_{{\hat x}}-u_{{\hat x}})$, and $l_{{\hat x}}$, $u_{{\hat x}}$ are the lower and upper bounds on ${\hat x}$. We tested these bounds in this paper, but we found their effect was highly similar to \eqref{eq: ReLU_MILPc}-\eqref{eq: ReLU_MILPd}; they were thus not utilized. The convex outer approximation of a ReLU, utilized in~\cite{Wong:2018}, was also tested and not utilized, since it provided no tightening advantage.

\begin{comment}
{\section{}\label{AppB}}
%In this section, we add normalization layers to the neural network model, and we define the associated system of linear inequalities utilized in this paper. 
We denote the training data input means and standard deviations as $\mu_{{\rm in}}={\rm mean}\{x_d\}$ and  $\bm{\Sigma}_{{\rm in}} ={\rm std}\{x_{d}\}$; similarly, we define $\mu_{{\rm out}}$ and $\bm{\Sigma}_{{\rm out}}$ as the output means and standard deviations associated with output shifted output $y_{d}-(\bm{J}^{\star}x_{d}+r^{\star})$. The full mapping from $x$ to output $y$ is given by 
\begin{align}
\tilde{x} & =\bm{\Sigma}_{{\rm in}}^{-1}\left(x-\mu_{{\rm in}}\right)\tag{normalize input}\\
\tilde{y} & =\bm{W}^{(1)}\sigma(\bm{W}^{(0)}\tilde{x}+b^{(0)})+b^{(1)}\tag{neural network}\\
y & =\bm{\Sigma}_{{\rm out}}\tilde{y}+\mu_{{\rm out}}+\bm{J}^{\star}x+r^{\star}.\tag{de-normalize output}
\end{align}
The explicit construction of the linear inequality $\bm{A}u-b\ge0$ is given on the GitHub page associated with this paper:~\cite{Chevalier_github:2024}.
\end{comment}
% which bounds the input via \eqref{eq: bound_input}, the output via \eqref{eq: bound_output}, the NN transformation via the normalized counterpart of \eqref{eq: relu_to_MILP}, and the binary relaxation via $0\le \beta \le 1$. Defining 

\bibliographystyle{IEEEtran}
\bibliography{references}

\end{document}